\documentclass[twocolumn]{emulateapj}
\usepackage{natbib}
\usepackage{amsmath}
\usepackage{longtable}
\usepackage{hyperref}
\usepackage{threeparttable}
\usepackage{array}

\begin{document}
\title{Defining Photometric Peculiar Type Ia Supernovae}
\author{S.~Gonz\'{a}lez-Gait\'{a}n$^{1,2}$}
\author{E.~Y.~Hsiao$^{3}$}
\author{G.~Pignata$^{4,1}$}
\author{F.~F\"orster$^{5,1}$}
\author{C.~P.~Guti\'errez$^{1,2}$}
\author{F.~Bufano$^{1,4}$}
\author{L.~Galbany$^{1,2}$}
\author{G.~Folatelli$^{6}$}
\author{M.~M.~Phillips$^{3}$}
\author{M.~Hamuy$^{2,1}$}
\author{J.~P.~Anderson$^{7}$}
\author{T.~de~Jaeger$^{1,2}$}

\affiliation{$^1$Millennium Institute of Astrophysics, Casilla 36-D, Santiago, Chile}
\affiliation{$^{2}$Departamento de Astronom\'ia, Universidad de Chile, Camino El Observatorio 1515, Las Condes, Santiago, Chile}
\affiliation{$^{3}$Carnegie Observatories, Las Campanas Observatory, Casilla 601, La Serena, Chile}
\affiliation{$^{4}$Departamento de Ciencias F\'isicas, Universidad Andres Bello, Avda. Republica 252, Santiago, Chile}
\affiliation{$^{5}$Centro de Modelamiento Matem\'atico, Universidad de Chile, Av. Blanco Encalada 2120 Piso 7, Santiago, Chile}
\affiliation{$^{6}$Kavli Institute for the Physics and Mathematics of the Universe, the University of Tokyo, Kashiwa, Japan 277-8583 (Kavli IPMU, WPI)}
\affiliation{$^{7}$European Southern Observatory, Alonso de C\'ordova 3107,
Casilla 19, Santiago, Chile}

\email{sgonzale@das.uchile.cl}

\begin{abstract}
We present a new photometric identification technique for SN~1991bg-like type Ia supernovae (SNe~Ia), i.e. objects with light-curve characteristics such as later primary maxima and absence of secondary peak in redder filters. This method is capable of selecting out this sub-group from the normal type Ia population. Furthermore, we find that recently identified peculiar sub-types such as SNe~Iax and super-Chandrasekhar SNe~Ia have similar photometric characteristics as 91bg-like SNe~Ia, namely the absence of secondary maxima and shoulders at longer wavelengths, and can also be classified with our technique. The similarity of these different SN~Ia sub-groups perhaps suggests common physical conditions. This typing methodology permits the photometric identification of peculiar SNe~Ia in large up-coming wide field surveys either to study them further or to obtain a pure sample of normal SNe~Ia for cosmological studies.
\end{abstract}


\section{Introduction}\label{intro}


Despite great progress over recent years from the observational as well as from the theoretical perspective, the nature of type Ia supernovae (SNe~Ia) remains elusive and under active debate. These luminous explosions have been used to obtain cosmological parameters with increasing precision \citep{Riess98,Perlmutter99,Conley11,Suzuki12}, yet the progenitor system remains still unknown. A key to unveil their origin relies in the study and understanding of the intrinsic diversity seen in their light-curves and colors, whose cosmic evolution ultimately also affects cosmology through systematics.

SNe~Ia constitute in fact a heterogeneous sample with differing light-curves whose primary variety is well explained by the relations of luminosity and light-curve shape \citep{Phillips93}, and of luminosity and color \citep{Riess96,Tripp98}. The variation of light-curve shape and luminosity is attributed to the amount of $^{56}$Ni produced in thermonuclear runaway of the exploding carbon-oxygen white dwarf (CO-WD) \citep{Mazzali07}. The nature of the companion star of the binary system and the physics of the explosion are still unknown. Models include single degenerate scenarios with a non-degenerate companion and the WD exploding near the Chandrasekhar mass \citep[e.g.][]{Nomoto84,Hachisu96,Dessart13} or at sub-Chandrasekhar mass \citep{Nomoto82b,Sim10,Kromer10}, as well as double degenerate scenarios with two WDs slowly coalescing \citep{Loren09,Shen12} or merging in a violent fashion \citep{Iben84,Pakmor12,Kromer13}; and even more exotic channels \citep{Kashi11,Wheeler12}. See \citet{Maoz13} for a review.

Extreme cases and outliers in the SN~Ia population were recognized early on: overluminous SNe~Ia or SN~1991T-like objects \citep[e.g][]{Filippenko92a,Phillips92,Maza94}, and subluminous or SN~1991bg-like ones \citep[e.g.][]{Filippenko92b,Leibundgut93,Hamuy94}. These objects make up for a considerable fraction of SNe Ia \citep[$\sim40\%$]{Li11b,Smartt09} and represent a key ingredient to understand the SN~Ia mechanism. Furthermore, in recent years new interesting outliers have been challenging further the general SN~Ia picture adding more heterogeneity and complexity: peculiar 2002cx-like SNe or ``SNe~Iax'' \citep[e.g.][]{Li03,Foley13x}, super-Chandrasekhar-mass candidates \cite[e.g.][]{Howell06,Scalzo12} and some peculiar SNe with few representatives in their groups: SN~2000cx-like \citep{Li01b,Candia03,Silverman13}, SN~2006bt \citep{Foley10} and PTF~10ops-like with 91bg-like SN~Ia characteristics yet with wide light-curve \citep{Maguire11,Kromer13}. Also, there is recent evidence for SNe~Ia with possible circumstellar material (CSM) interaction \citep{Hamuy03,Dilday12,Silverman13csm}. Finally, new intriguing objects borderline of thermonuclear explosions are emerging: the Calcium-rich SN group \citep[Ca-rich:][]{Perets10,Perets11b,Valenti13,Kasliwal12} and extremely underluminous ``SN.Ia'' candidates (\citealt{Kasliwal10} although see \citealt{Drout13,Drout14}).


This plethora of thermonuclear events asks for an urgent classification scheme and hopefully a proper theoretical explanation of the diversity of observables that will link them together to possible progenitor channels. The investigation of the characteristics of all these objects can help us carve the way. The method presented in this paper is such an attempt based solely on photometry and light-curve analysis of a large public SN~Ia dataset. 

The original aim of the present work was to separate SN~1991bg SNe~Ia from the normal population in an efficient purely photometric fashion. Several studies of 91bg-like SNe~Ia have been undertaken in the past \citep{Garnavich04,Taubenberger08,Kasliwal08}. These SNe are known to be fainter with $\mathrm{mag}(B)\gtrsim-18$ at maximum and to have a faster decline of the light-curve than their normal counterpart. Such a behaviour has been best portrayed by light-curve parameters such as $\Delta m_{15}$ and stretch from different fitters \citep[e.g.][]{Hamuy96b,Riess96,Jha07,Perlmutter97,Guy07}, always putting 91bg-like SNe~Ia at the faint and fast extreme of the light-curve width-luminosity relation with $\mathrm{stretch}\lesssim0.75$ or $\Delta m_{15}(B)\gtrsim1.7$. In addition, 91bg-like events are much cooler as seen through their spectra with presence of strong \ion{Ti}{2} lines and their red colors, $(B-V)_{B_{\mathrm{max}}}\gtrsim0.3$. Such a deviation from the normal population that has made them unsuitable for cosmology. The different color evolution is particularly striking in the redder optical bands and the near-infrared (NIR) where neither shoulder nor secondary maxima are observed as opposed to classical SNe~Ia \citep{Phillips11}. The color evolution presents also other less noticeable differences in the relative time of maximum light in the different bands: 91bg-like SNe~Ia have maxima in redder band that occur after the maximum in bluer bands, whereas for normal events it happens before. These different features can been explained with the differing spectroscopic and color evolution of SNe~Ia \citep[e.g.][]{Kasen07}: 91bg-like objects are cooler, therefore redder, and the recombination temperature of \ion{Fe}{3} to \ion{Fe}{2} for them happens earlier. The onset of this recombination determines the red and NIR secondary maxima since the \ion{Fe}{2} line blanketing absorbs flux in the blue that is re-emitted at longer wavelengths.

The 91bg-like photometric typing technique we present here reveals surprising adequacy to adjust other peculiar SNe~Ia that have atypical light-curves such as SNe~Iax, SN~2006bt-like and super-Chandrasekhar SNe~Ia, and the possibility to photometrically classify them as well. In the current and coming age of large wide field surveys such as the Dark Enery Survey \citep[DES,][]{Sako11}, the Large Synoptic Survey Telecope \citep[LSST,][]{Ivezic11}, where thousands of transients are routinely discovered and the spectroscopic follow-up becomes expensive, the need of a photometric identification technique that provides a way to classify and recognize different sub-groups of SNe~Ia for different scientific studies becomes imperative.

The paper is organized as follows: in section~\ref{analysis}, we present the data, as well as our photometric typing technique based on light-curve fits, validating it with different light-curve and spectroscopic diagnostics. Section~\ref{results} presents the results of the classification for different sub-samples of SNe~Ia and in section~\ref{discussion} we investigate contamination from core-collapse SNe, the relevance and impact of the training sample used in the classification, and finally we look for possible common physical origins of the similar photometric SN~Ia groups. We summarize in section~\ref{summary}.


\section{Analysis}\label{analysis}

\subsection{Data}
In this work, we make use of several large low-redshift ($z<0.1$) SN~Ia samples from the literature. Multi-band photometry is available for more than 500 SNe~Ia obtained through the effort of several teams, including the Cal\'an/Tololo survey \citep{Hamuy96c}, the Carnegie Supernova Project CSP \citep{Contreras10,Stritzinger11}, Center for Astrophysics CfA \citep{Hicken09a,Hicken12}, Lick Observatory Supernova Search \citep{Ganeshalingam10} and many more. Besides analyzing SN photometry obtained exclusively after 1980, initially we do no restrict on the source of the photometry used; as long as it is internally consistent, we do not require the type of precise calibration needed for cosmology.
In order to test the typing technique, we also use a collection of nearby core-collapse (CC) SNe from the literature. A summary of the number of SNe before and after the different light-curve cuts we will apply is shown in table~\ref{fomtable}. A list of SNe~Ia used and their sources is presented in table~\ref{Iaphot} and of CC~SNe in table~\ref{CCphot}.

\subsection{Methodology: light-curve fits}

\begin{figure*}[htbp]
\centering
\includegraphics[width=0.49\linewidth]{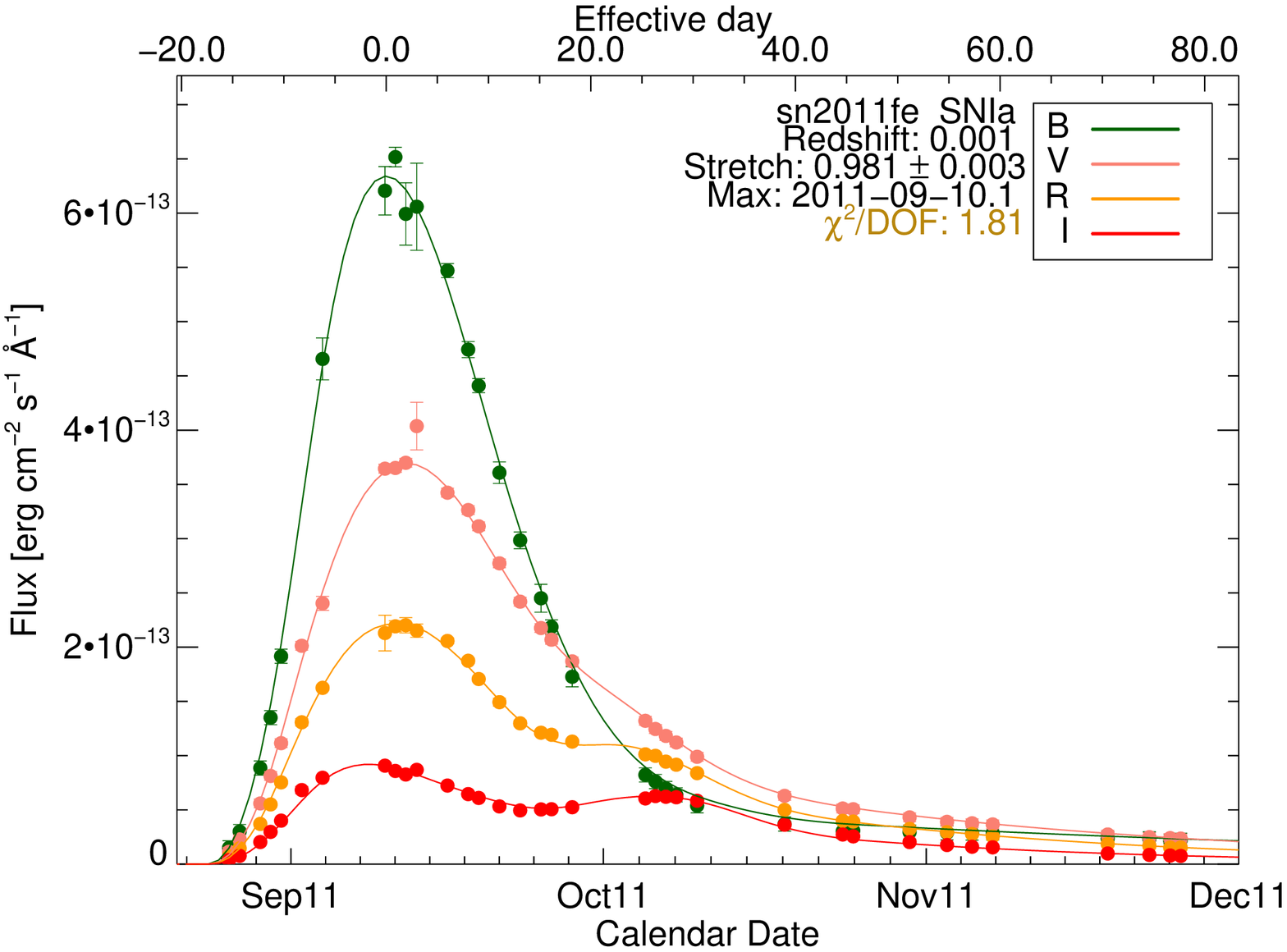}
\includegraphics[width=0.49\linewidth]{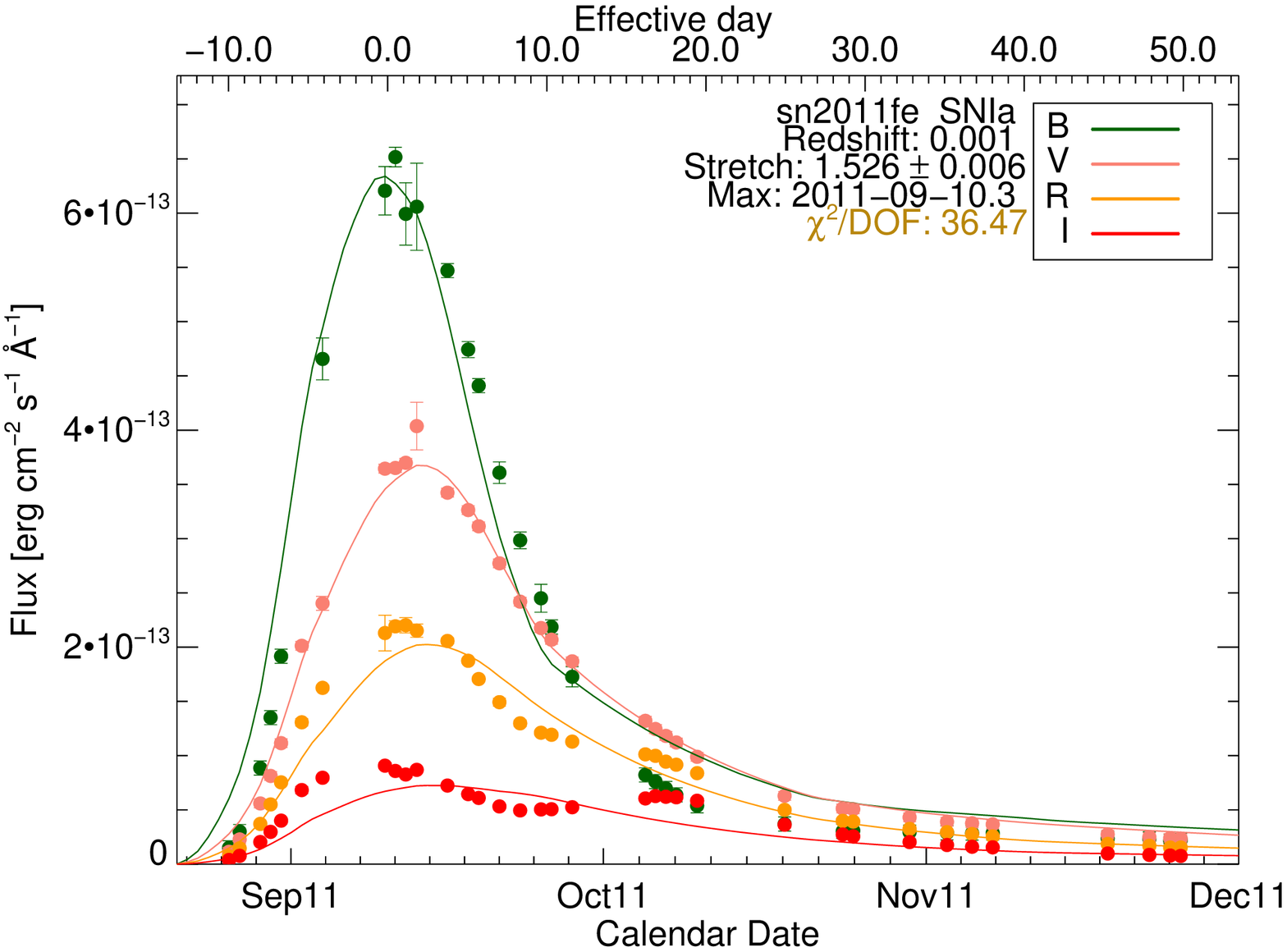}
\caption{Example SiFTO light-curve fits of SNe~Ia with better normal (\emph{left}) than 91bg-like (\emph{right}) Ia template fits. SN~2011fe has $s_{\mathrm{Ia}}=0.98$, $\chi^2_{\nu}$(Ia)$=1.8$ and $\chi^2_{\nu}$(91bg)$=36.5$.}
        \label{lcplots1}
\end{figure*}

\begin{figure*}[htbp]
\centering
\includegraphics[width=0.49\linewidth]{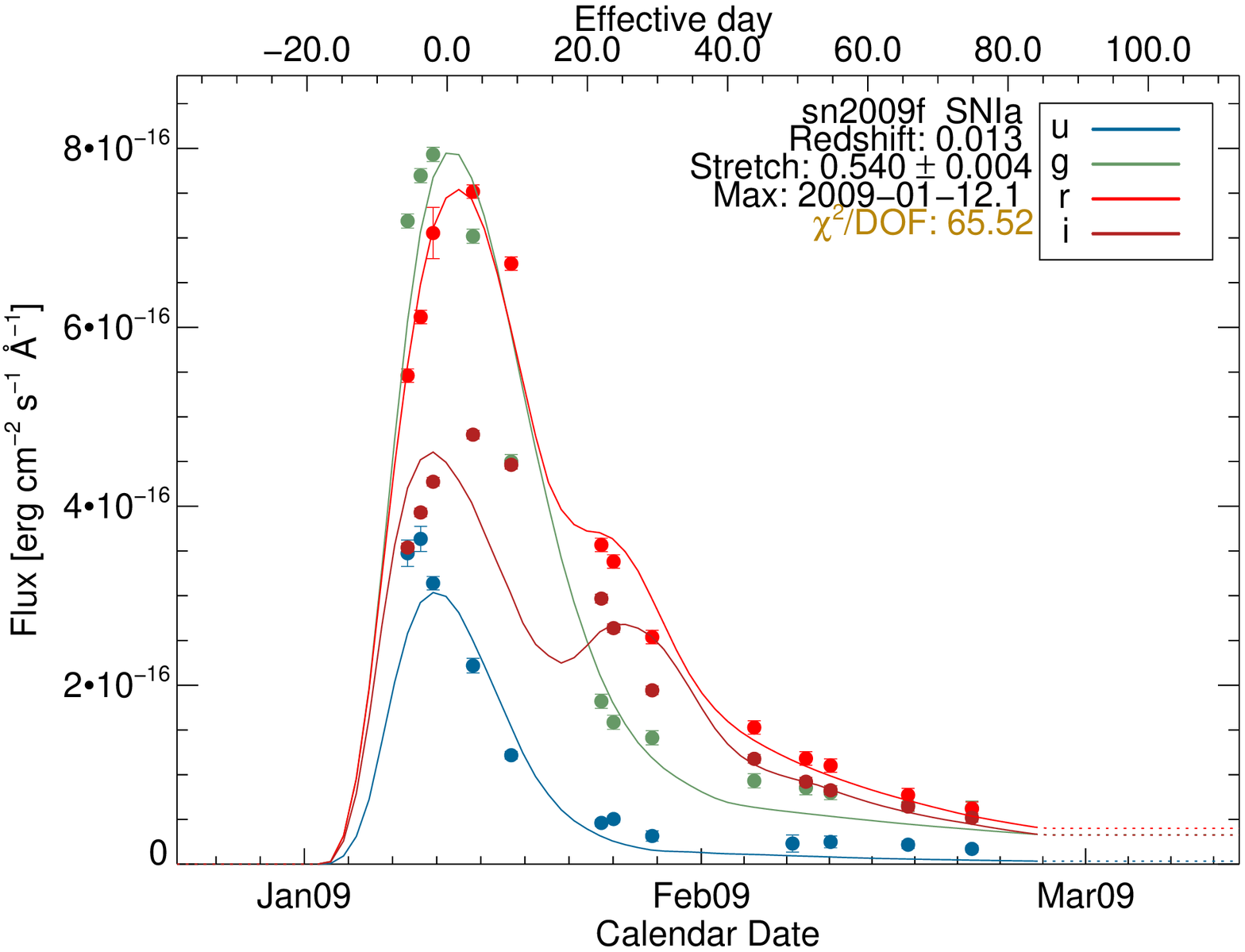}
\includegraphics[width=0.49\linewidth]{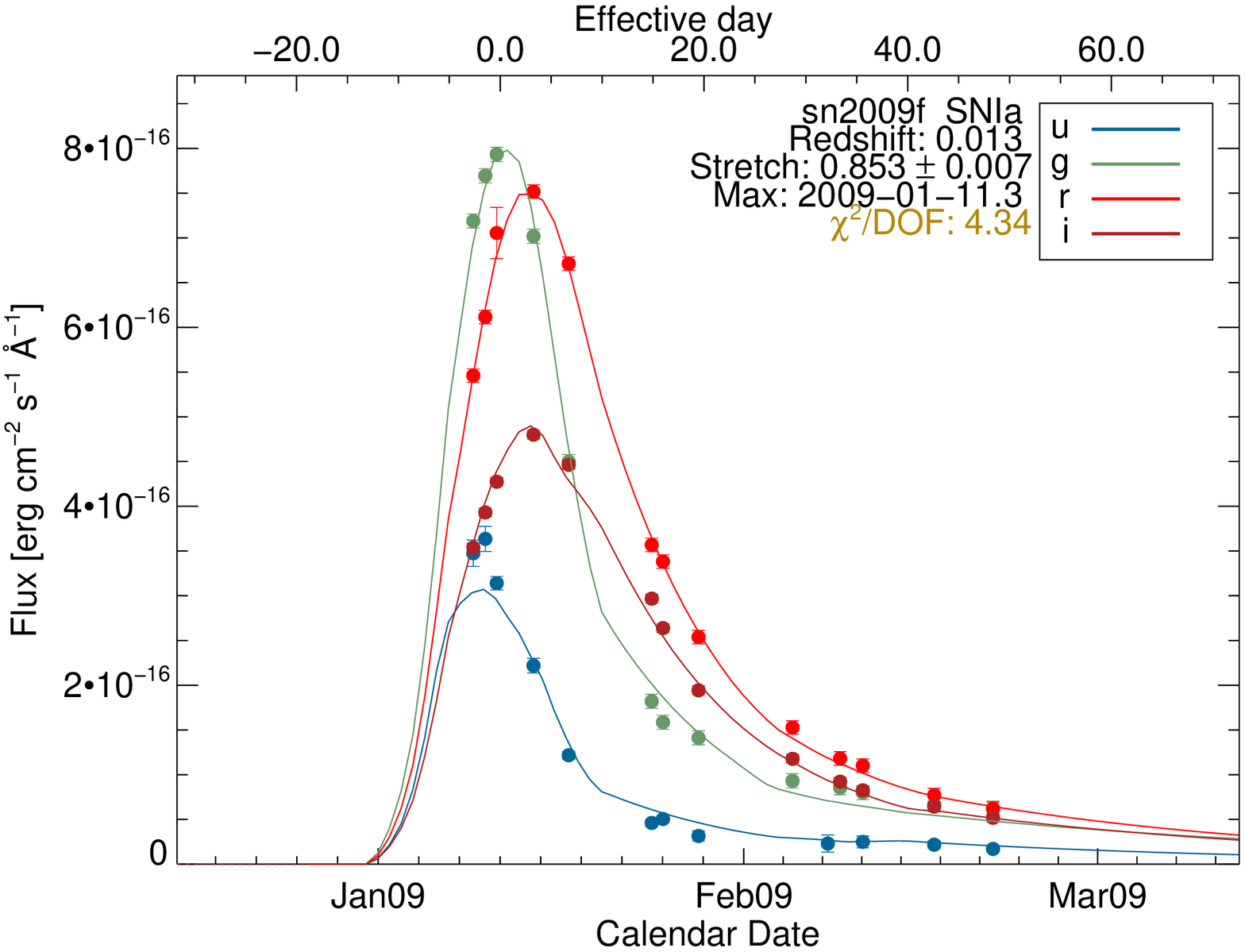}
\caption{Example SiFTO light-curve fits with worse normal (\emph{left}) than 91bg-like (\emph{right}) Ia template fits. SN~2009F has $s_{\mathrm{Ia}}=0.54$ ($s_{\mathrm{91bg}}=0.85$), $\chi^2_{\nu}$(Ia)$=65.6$ and $\chi^2_{\nu}$(91bg)$=4.3$.} 
        \label{lcplots2}
\end{figure*}


We use here SiFTO \citep{Conley08} to fit the optical light-curves of all SNe~Ia with two different templates that reproduce the aforementioned differences between normal and 91bg-like SNe~Ia. SiFTO is a powerful and versatile light-curve fitter that manipulates spectral templates to simultaneously match the multi-band photometric data of a particular SN with heliocentric redshift $z_{\mathrm{hel}}$. The spectral template at a given epoch $t$ is adjusted to the observer-frame fluxes through multiplicative factors $n_f$. A stretch parameterization $s$ is used to describe the shape of the light-curve and is defined as a factor multiplying the time axis \citep{Perlmutter97,Goldhaber01}. The time of peak luminosity in $B$, $t_0$, is an additional fit parameter. The light-curve fit thus minimizes following $\chi^2$ function:

\begin{equation}
\chi^2=\sum_{j=1}^{N_f}\sum_{i=1}^{N_j}\frac{\left[F_j(t,t_0,n_f,s,z_{\mathrm{hel}})-f_{ij}\right]^2}{\sigma_{ij}^{2}},
\end{equation}
where $N_j$ is the number of data points in the $j$th filter for $N_f$ filters, $f_{ij}$ are the observed fluxes with corresponding errors $\sigma_{ij}$ and $F_j$ is the modelled flux given by the integration of the SED through the given filter $j$ at a certain epoch $t$ and dependent on the model parameters. The $\chi^2$ is a general score of the fit taking all input bands together. The SiFTO color, $\mathcal{C}$, is obtained adjusting the SED to observed colors (via the normalization factors $n_i$) corrected only for Milky Way extinction with values from \citet{Schlafly11}. No correction for the host reddening is attempted due to our poor understanding of it.

Given its nature, SiFTO is highly dependent on the spectral template series. We use the one developed by \citet{Hsiao07} for normal SNe~Ia, commonly used for cosmology \citep{Conley11} and other SN~Ia studies \citep[e.g.][]{Pan13}, and the one by \citet{Nugent02}\footnote{\url{http://supernova.lbl.gov/~nugent/nugent_templates.html}} for 91bg-like SNe~Ia, previously used to identify objects with 91bg-like characteristics \citep{Gonzalez11,Maguire11,Sullivan11a}. The normal template was constructed with a large sample of SNe~Ia at low and high redshift, whereas the 91bg-like SN~Ia template comes from the classical subluminous SN 1991bg and SN 1999by. Although there are many more 91bg-like SNe~Ia nowadays to improve this template, we show in the following that it is appropriate for our purposes.

In order to ensure a proper fit, we require at least two filters, each with at least one data point between -15 and 0 days and one between 0 and 25 days past $B$-band maximum. We perform fits using all available optical photometry down to 85 days past maximum. If different photometric calibrations from different instruments exist per filter set, we choose only one set, based on the number of data points, to ensure consistent photometry per filter. Ideally, as will be shown later, red bands such as $R$ or $r$ and $I$ or $i$ are benefitial to fully exploit the range of possible SN~Ia sub-groups. The SiFTO fits are performed with both ``normal'' and ``91bg'' templates. Some example fits are shown in figure \ref{lcplots1} and figure~\ref{lcplots2}, where one can compare the quality of the fits for the two templates for two SNe~Ia: SN~2011fe and SN~2009F. The first presents better normal template fit as seen in the figure as well as on the lower overall reduced fit quality, $\chi^2_{\nu}$(Ia)$<\chi^2_{\nu}$(91bg), whereas the other shows better 91bg template fits and have $\chi^2_{\nu}$(Ia)$>\chi^2_{\nu}$(91bg). We first use this simple photometric criterion (criterion 1) to separate the SN~Ia population into photometric normal and SN~1991bg/SN~1999by-like objects. We note that for fits with a 91bg template, a ``stretch'' has a different meaning since it is with respect to a typical 91bg-like SN~Ia, so SN~1991bg and SN~1999by have a ``normal template'' stretch of $s_{\mathrm{Ia}}=0.46$ and $0.62$ but a ``91bg template'' stretch of $s_{\mathrm{91bg}}=0.86$ and $1.02$,respectively. The definition of standard $s=1$ SN is shifted in both cases. We also emphasize that the stretch parameter is a factor defined in $B$-band, so that it evaluates bests the variation near this wavelength.

\subsection{Blue- and red-band template fits}\label{bluered}

\begin{figure*}[htbp]
\centering
\includegraphics[width=1.0\linewidth]{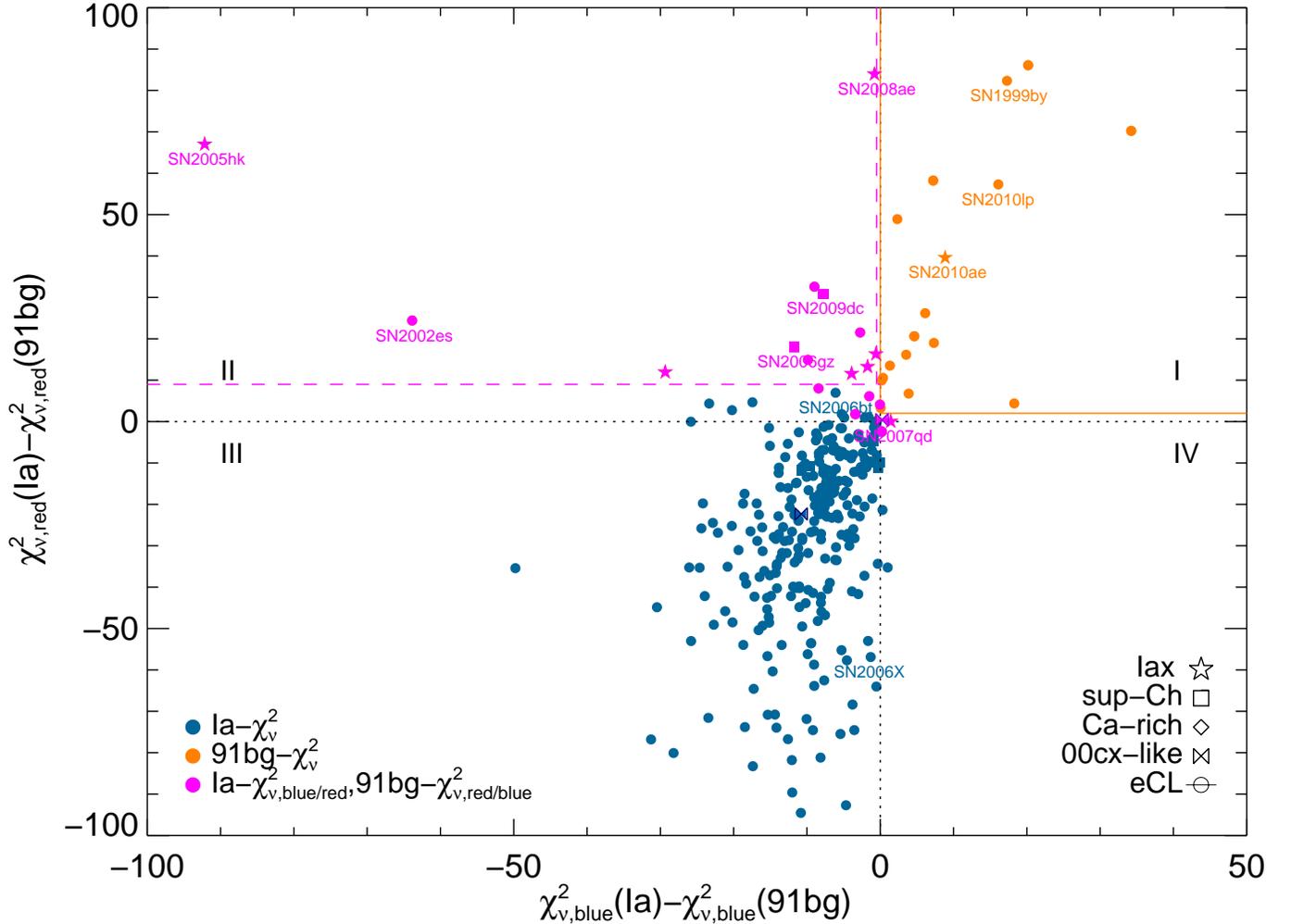}
\caption{Difference between normal and 91bg template fit qualities in the blue bands ($U/u$, $B$ and $g$): $\chi^2_{\nu,\mathrm{blue}}(\mathrm{Ia})-\chi^2_{\nu,\mathrm{blue}}(\mathrm{91bg})$, versus the difference between normal and 91bg template fit qualities in the red bands ($V$, $R/r$ and $I/i$): $\chi^2_{\nu,\mathrm{red}}(\mathrm{Ia})-\chi^2_{\nu,\mathrm{red}}(\mathrm{91bg})$. Zero $\chi^2_{\nu}$ difference dotted lines are shown, solid orange lines denote the maximum FoM(91bg) box (or quadrant I) of 91bg-like objects according to both criteria, overall and blue/red-band fits. The blue symbols represent normal photometric candidates, where the overall fits are consistent with a normal SN~Ia template. Orange symbols are 91bg-like candidates according to both criteria (overall, blue- and red-band) fits that are delimited by the orange box. Purple symbols are objects that are 91bg-like according to the overall fit (criterion 1) but not according to criterion 2 of blue/red-band fits, i.e. they lie outside of the orange box. A box of minimum FoM(pec) is shown as purple dashed lines, although it is not used as part of the classification. Different symbol shapes show known peculiar or extreme objects such as: SNe~Iax (stars), super-Chandrasekhar SNe~Ia (squares) and SN~2000cx-like objects (bowtie) and extremely cool SNe~Ia or ``eCL'' (circles with horizontal lines) as in \citet{Folatelli13}.}
       \label{brchisq}
\end{figure*}

Although the 91bg-like SN~Ia typing technique based on the overall $\chi^2_{\nu}$ comparison between normal and 91bg template fits is quite good at separating 91bg-like from normal SNe~Ia according to spectroscopic classifiers, as will be shown in next sections, there is a fraction of non 91bg-like objects that are typed as such. Investigating closely these objects, we find that many of them are actually peculiar SNe~Ia of a different kind. SNe~Iax are all typed as 91bg-like SNe~Ia, super-Chandrasekhar objects as well, and even SN~2013bh, the only other 2000cx-like member in the literature \citep{Silverman13} is also picked up. Although at first intriguing, one can understand this result as their light-curves in redder bands resemble those of typical 91bg-like objects more than those of normal ones, showing less or no shoulder and secondary maxima.

Nonetheless, one can also see that their photometric behaviour is different from the true 91bg-like objects in the bluer bands and that the overall $\chi^2_{\nu}$ is being driven by extremely poor normal template fits in the redder bands. To investigate this further, we re-fit all our sample restricting the fit to ``blue-bands'', i.e. using following available filters simultaneously: $u$ or $U$, $B$ and $g$, with both templates, normal and 91bg-like. Then, we do the same but restricting the fit to ``red-bands'', i.e. using simultaneously filters $V$, $r$ or $R$, and $i$ or $I$. We require for this at least one filter with one point prior to maximum and one after maximum. Figure~\ref{brchisq} shows the resulting difference in fit quality between normal and 91bg templates for the red-band versus blue-band fits. We denote four quadrants in this plot, which we will refer to throughout the paper. In this figure we can see that if we add a restriction to the photometric definition of a 91bg-like SN~Ia based on the result of the comparison of fits in the blue and red bands, i.e. $\chi^2_{\nu,\mathrm{blue}}(\mathrm{91bg}) < \chi^2_{\nu,\mathrm{blue}}(\mathrm{Ia})$ and $\chi^2_{\nu,\mathrm{red}}(\mathrm{91bg}) < \chi^2_{\nu,\mathrm{red}}(\mathrm{Ia})$ (vertical and horizontal dotted lines), criterion 2, we discard almost all peculiar SNe~Ia, in particular all SNe~Iax except for SN~2010ae and SN~2007qd, and all super-Chandrasekhar SNe~Ia, while keeping the known spectroscopic 91bg-like SNe~Ia. It is worth mentioning that since only two objects have better 91bg template fits in blue bands but worse in red bands (quadrant IV), this criterion could also only consist of just a blue-band cut and no red-band cut to include most objects.

The separation between these groups does not necessarily need to be at exactly $\chi^2_{\nu,\mathrm{blue}}(\mathrm{91bg})=\chi^2_{\nu,\mathrm{blue}}(\mathrm{Ia})$ and $\chi^2_{\nu,\mathrm{red}}(\mathrm{91bg})=\chi^2_{\nu,\mathrm{red}}(\mathrm{Ia})$ (dotted lines in fiure~\ref{brchisq}), and could instead be offset from these lines. To investigate this further, we use a common photometric typing diagnostic \citep[e.g.][]{Kessler10}, the Figure of Merit (FoM), given by the efficiency of the classification ($\epsilon$), i.e. the fraction objects of a given type correctly tagged, and the purity ($P$) or fraction of classified objects that really are of that type, i.e. a measure of the false positive tags:

\begin{eqnarray} \label{fom}
\mathrm{FoM}&=&\epsilon\times P \\ 
&=&\frac{N^{\mathrm{true}}}{N^{\mathrm{tot}}}\times\frac{N^{\mathrm{true}}}{N^{\mathrm{true}}+N^{\mathrm{false}}}, \nonumber
\end{eqnarray}
where $N_{\mathrm{true}}$ is the number of correctly identified objects of a given type (e.g. 91bg-like objects), $N_{\mathrm{tot}}$ is the total input number of that type and $N_{\mathrm{false}}$ is the number of objects falsely tagged as objects of that given type.

Calculating this FoM for known 91bg-like objects, FoM(91bg), for different boxes around $\chi^2_{\nu,\mathrm{blue}}(\mathrm{91bg})-\chi^2_{\nu,\mathrm{blue}}(\mathrm{Ia})=\pm20$ and $\chi^2_{\nu,\mathrm{red}}(\mathrm{91bg})-\chi^2_{\nu,\mathrm{red}}(\mathrm{Ia})=\pm20$ and looking for the maximum FoM, we find the areas highlighted in orange in figure~\ref{brchisq}. This box is similar to the original $\Delta\chi^2_{\nu}=0$ with a slight vertical offset of  $\chi^2_{\nu,\mathrm{red}}(\mathrm{Ia})-\chi^2_{\nu,\mathrm{red}}(\mathrm{91bg})=2$, which leaves out peculiar SN~2007qd and SN~2013bh. Objects inside this region, besides having a better overall 91bg template fit, are also 91bg-like candidates according to the blue/red-band fits (criterion 1 and 2). This selected group is located in quadrant I of the figure and symbols therein are shown in orange. SN~1991bg-like candidates according to the overall $\chi^2_{\nu}(\mathrm{91bg})$ (criterion 1) that are not in this region, i.e. do not fulfill criterion 2, are denoted with purple symbols in the figure (mostly in quadrant II). Blue symbols are normal candidates according to the overall fit. 

Alternatively, to eliminate the first criterion and be able to select peculiar SNe~Ia, i.e. those objects in purple, one can also find a region of maximum FoM(pec) in a similar fashion to FoM(91bg). This results in the area shown in purple dashed lines. This zone however does not include the two objects in quadrant IV, SN~2007qd and SN~2013bh. These dividing lines are sensitive to the training sample used and will therefore be studied in more detail in section~\ref{discussion}. 

\subsection{Comparison with other light-curve parameters}

\begin{figure*}
\centering
\includegraphics[width=0.8\linewidth]{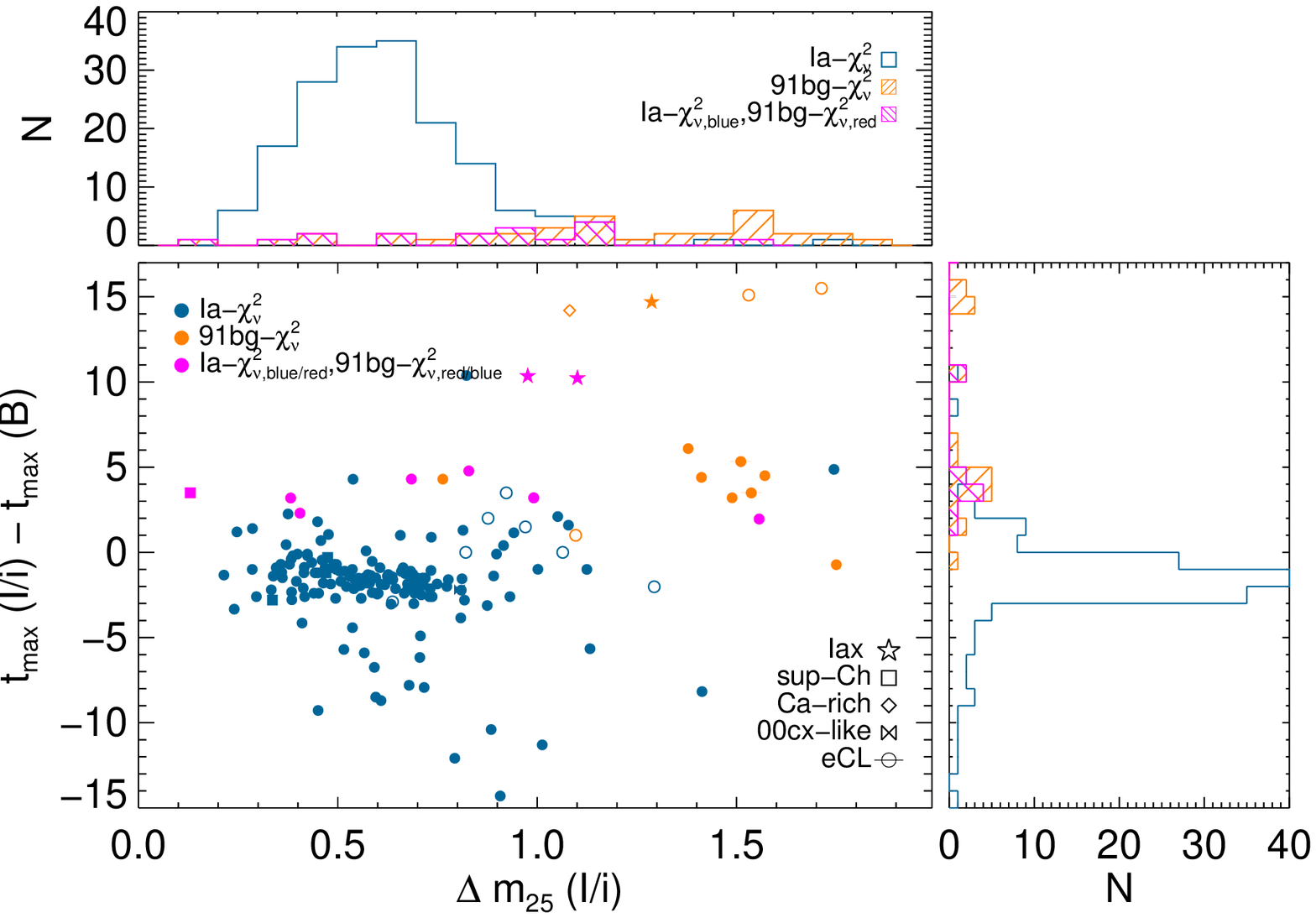}
\caption{Difference between day of maximum light in $I$ or $i$-band and maximum in $B$-band versus $\Delta m_{25}$ in $I$ or $i$-band. SNe with overall fits more consistent with normal template fits are shown as filled blue symbols. Objects that have better overall 91bg template fits (fufilling criterion 1) and better blue/red-band 91bg template fits (fulfilling criterion 2) are shown in orange, whereas those with better overall 91bg template fits yet better blue/red-band normal template fits are shown in purple. Symbols represent known peculiar or extreme objects: Iax (stars), super-Chandra (square), Ca-rich (diamond), 00cx-like (bowtie) and eCL (circles with line). Additionally, objects for which not enough data was available to check blue- and red-band fits are shown as open symbols for photometric normal (\emph{blue}) and 91bg-like (\emph{orange}) SNe~Ia. Upper and right panels show the distributions for both samples in empty blue and filled orange boxes respectively.}
       \label{dm25max}
\end{figure*}


In the examples shown in figures \ref{lcplots1} and \ref{lcplots2}, it seems evident that the shoulder and secondary maxima of bands at longer wavelength determine the quality of the fits since they are present in normal SNe~Ia as opposed to 91bg-like objects. Nevertheless, it is important noting that the time of maximum in each band also plays a crucial role. To test these light-curve characteristics independently from SiFTO, we calculate the time of maximum in each band using a fourth degree polynomial fit and we also look for a secondary maximum with another polynomial fit. If the secondary maximum is not very separated in time from the primary maximum in redder bands such as $R$/$r$ or $I$/$i$, we are not able to define such a secondary; in fact, we find a secondary maximum in $I$/$i$ for only 2 (of 48) 91bg-like candidates according to criterium 1, whereas this is found for more than 90 (of 295) normal SNe~Ia. A more continuous parameter that can be defined for a larger sample is the magnitude change after 25 days past maximum or $\Delta m_{25}$: if a SN has a secondary shoulder or a maximum that typically occurs 20-35 days past maximum for redder bands, then this parameter should somewhat reflect this. In figure \ref{dm25max} we show two light-curve diagnostics, the difference in maximum between two different bands, $t_{\max}(B)-t_{\max}(I/i)$ and $\Delta m_{25}(I/i)$ for the samples selected with the mechanism shown in the previous section. Clearly, as expected, the normal SNe~Ia according to our light-curve fits have maxima in red bands that happen earlier and they also have lower $\Delta m_{25}(I/i)$ than 91bg-like candidates that had better overall, blue and red band 91bg template fits (filled orange). Objects with better overall but worse blue/red 91bg template fits (filled purple) also differ from normal SN~Ia candidates, although not as strongly as 91bg-like candidates. These trends happen in $R$/$r$, are stronger in $I$/$i$ and are exacerbated in the NIR \citep{Phillips11}.

\subsection{Comparison with spectral classificators}

\begin{figure}[htbp]
\centering
\includegraphics[width=1.\linewidth]{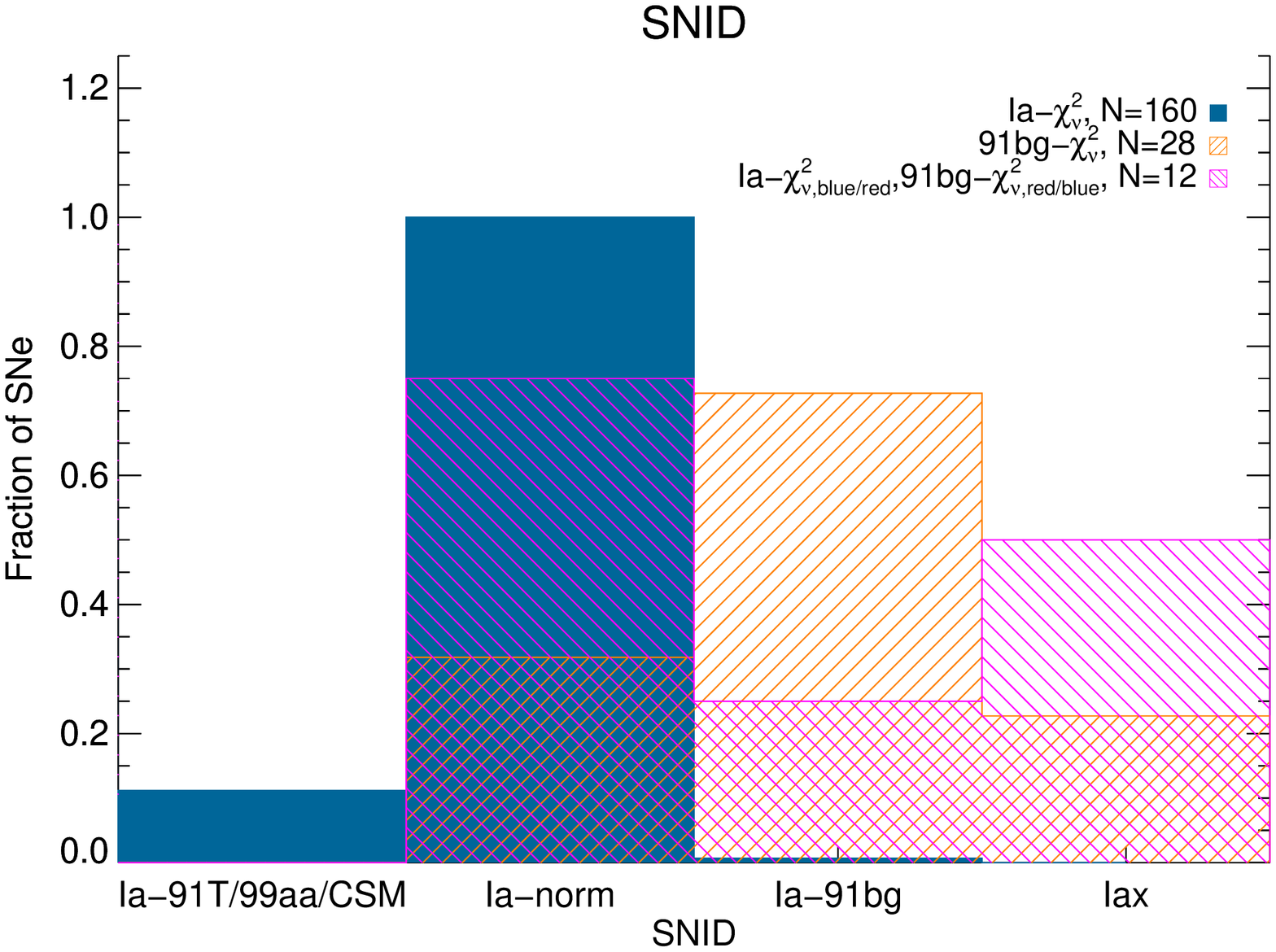}
\includegraphics[width=1.\linewidth]{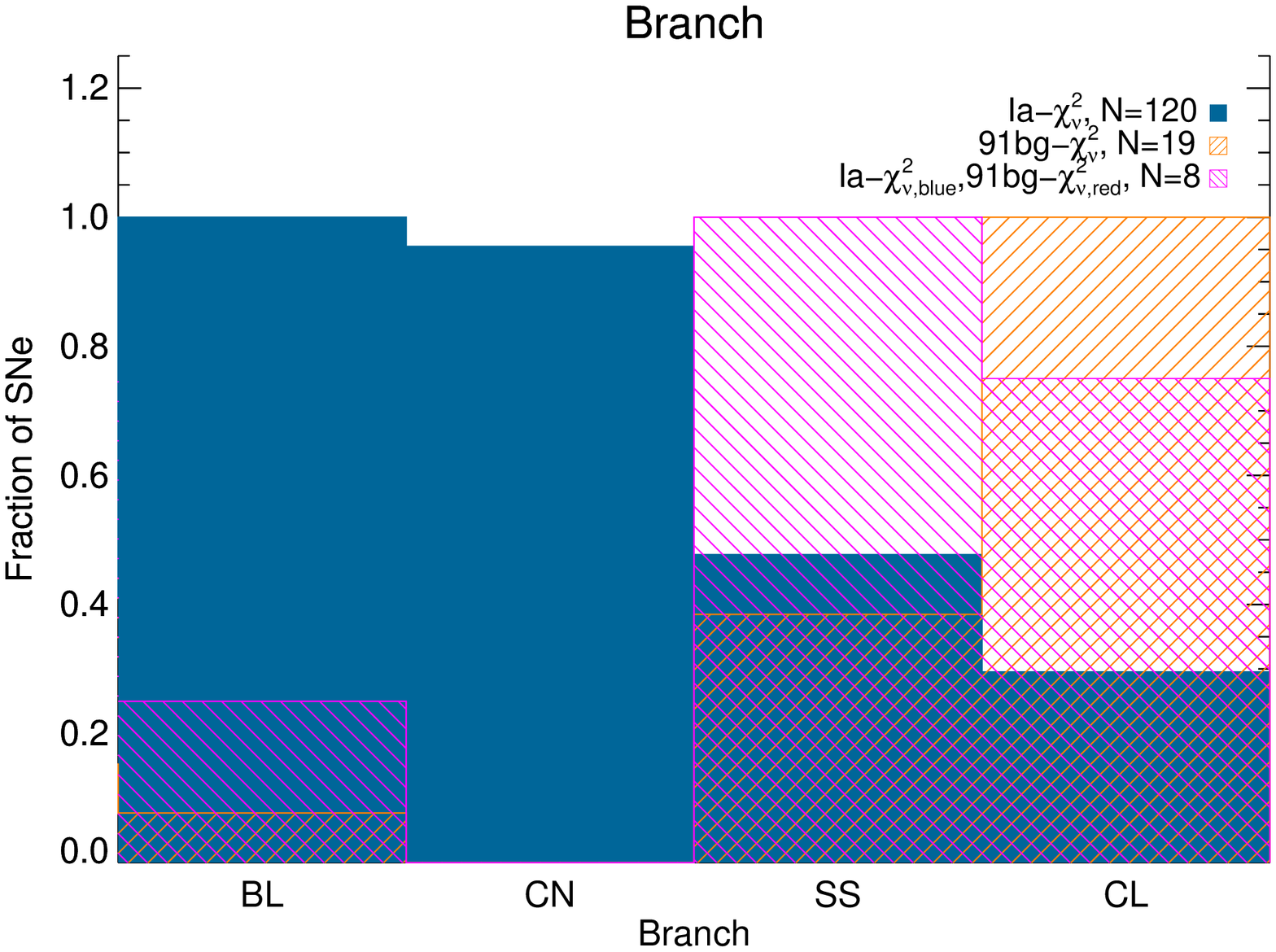}
\includegraphics[width=1.\linewidth]{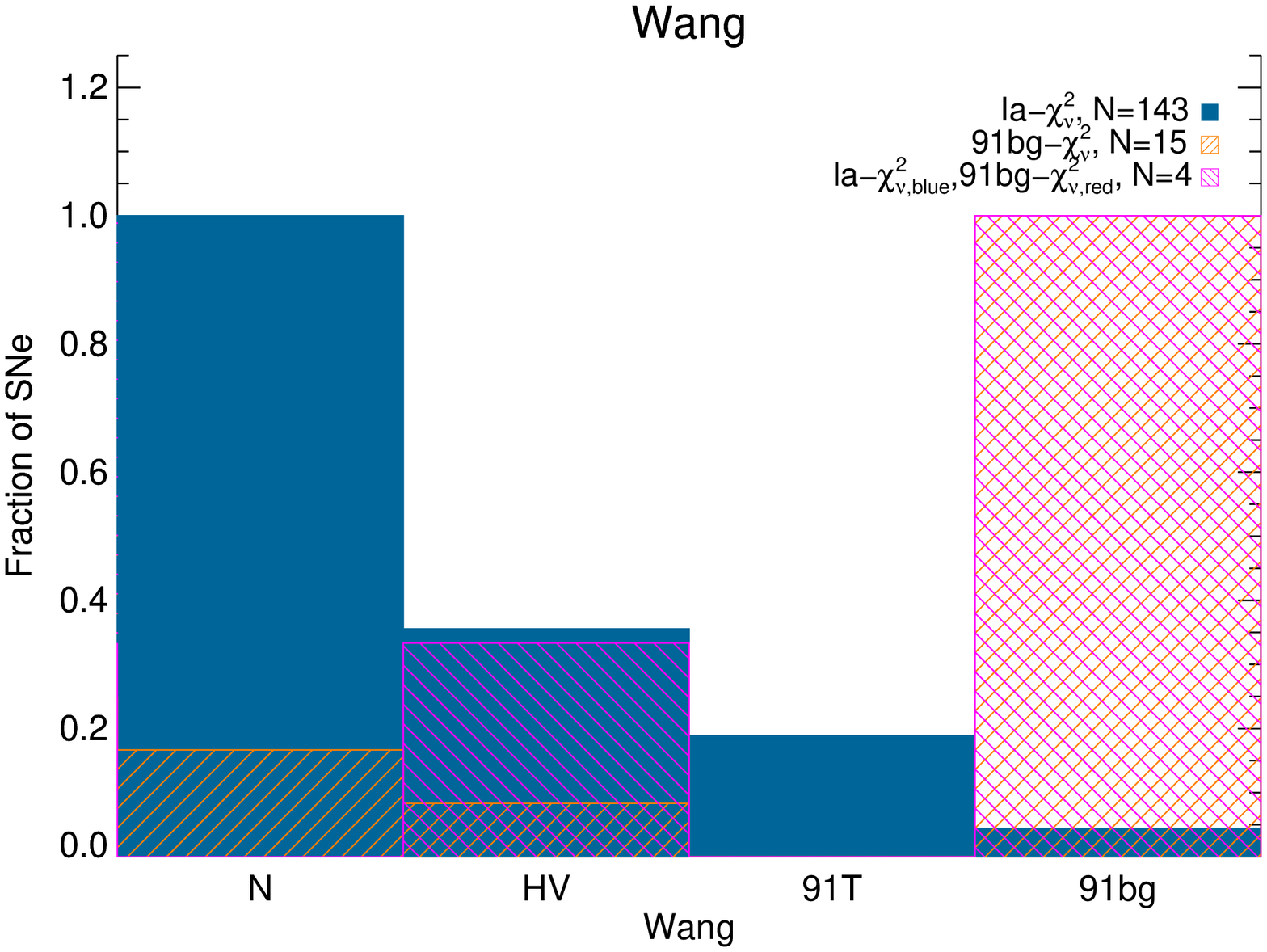}
\caption{Comparison of the photometric classification technique to detect normal, 91bg-like and other peculiar SNe~Ia with spectral classifiers: a) SNID \citep{Blondin07}, b) \citet{Branch06} and c) \citet{Wang09b}. Blue histograms (open) show objects with better overall normal template fit, orange (filled with $45^{\circ}$ lines) histograms show objects with better overall 91bg template fit and better blue/red-band 91bg template fit (criterion 1 and 2), wherease purple (filled with $-45^{\circ}$ lines) are objects with better overall 91bg template fit but worse blue or red-band 91bg template fit (criterion 1). Histograms are normalized to one and the total number for each histogram is indicated in the figures.}
       \label{speclass}
\end{figure}

The standard classification of supernovae is defined by identifying spectral features near maximum light. Several classification schemes for SNe~Ia have been proposed, notably those by \citet{Benetti05}, \citet{Branch06} and \citet{Wang09b}, all based on prominent line properties such as Silicon strength, velocity and velocity gradient. Additionally, some automated classification tools such as SNID \citep{Blondin07} or GELATO\footnote{Available at \url{https://gelato.tng.iac.es/login.cg}} \citep{Harutyunyan08} have been presented for general use.  

Using a large sample of classifications from SNID given by \citet{Silverman12}, we compare them to our photometric classificator. 220 SNe~Ia match their sample and we show the result in figure~\ref{speclass}. SNID gives the following classifications for SNe~Ia: Ia-norm, Ia-91T, Ia-91bg, Ia-CSM, Ia-99aa and Ia-02cx. We group here Ia-91T, Ia-99aa and Ia-CSM together into a single bin that, as can be seen, has no further influence in our classification. From this figure, one can see that most of the 187 objects ($99.5\%$) photometrically classifed as normal SNe~Ia are also normal according to SNID. On the other hand, we do see that photometric 91bg-like candidates are found in other bins. In particular, 7 objects leak into the Ia-norm bin. Some of these objects, like SN 2007on, are borderline between photometric normal and 91bg-like according to our overall fits having very similar $\chi^2_{\nu}$, but some others have quite interesting properties that make them stand out, like super-Chandrasekhar SNe~Ia. Most of these can be identified with quality of fit in particular filter sets, as shown in last section, and shown in the figure with purple histograms. All SNe~Iax are typed as photometric 91bg-like SNe~Ia according to the overall fit. This result is possibly physically interesting per se, and for cosmological studies seeking for purely normal SNe~Ia, it also presents a way to photometrically take out more than one group of peculiar objects simultaneously. For a 91bg-like SN~Ia typing technique however, this could be worrisome. Fortunately, the inclusion of blue/red-band fits permits us to discriminate quite well between both groups as well.


\citet{Blondin12} also present a large spectroscopic dataset and publish the classifications of \citet{Branch06} and \citet{Wang09b} for their sample. We find 189 matches and show the comparison in figure~\ref{speclass} for these spectral groups. In general we find agreement with our typing (the ``Cool'', CL, group corresponds to the 91bg-like for \citealt{Branch06}). We again find that some objects that we type photometrically as 91bg-like according to the overall fit are misidentified according to these spectral classifications. Nevertheless, most of these actually correspond to SNe~Iax and super-Chandrasekhar SNe~Ia and are mostly included in the purple histograms that take into account blue/red-band fits. Thus, from this we see that the technique here presented allows to differentiate quite successfully normal, 91bg-like and other peculiar SNe~Ia, notably SNe~Iax. More on these other sub-classes will be discussed in next sections.

In general, we find a remarkable agreement between our photometric typing and the different spectroscopic classifiers. This is impressive given that SiFTO was not devised originally for this purpose and it opens up promising possibilities for other transient surveys.





\section{Results}\label{results}
\begin{figure*}[htbp]
\centering
\includegraphics[width=0.75\linewidth]{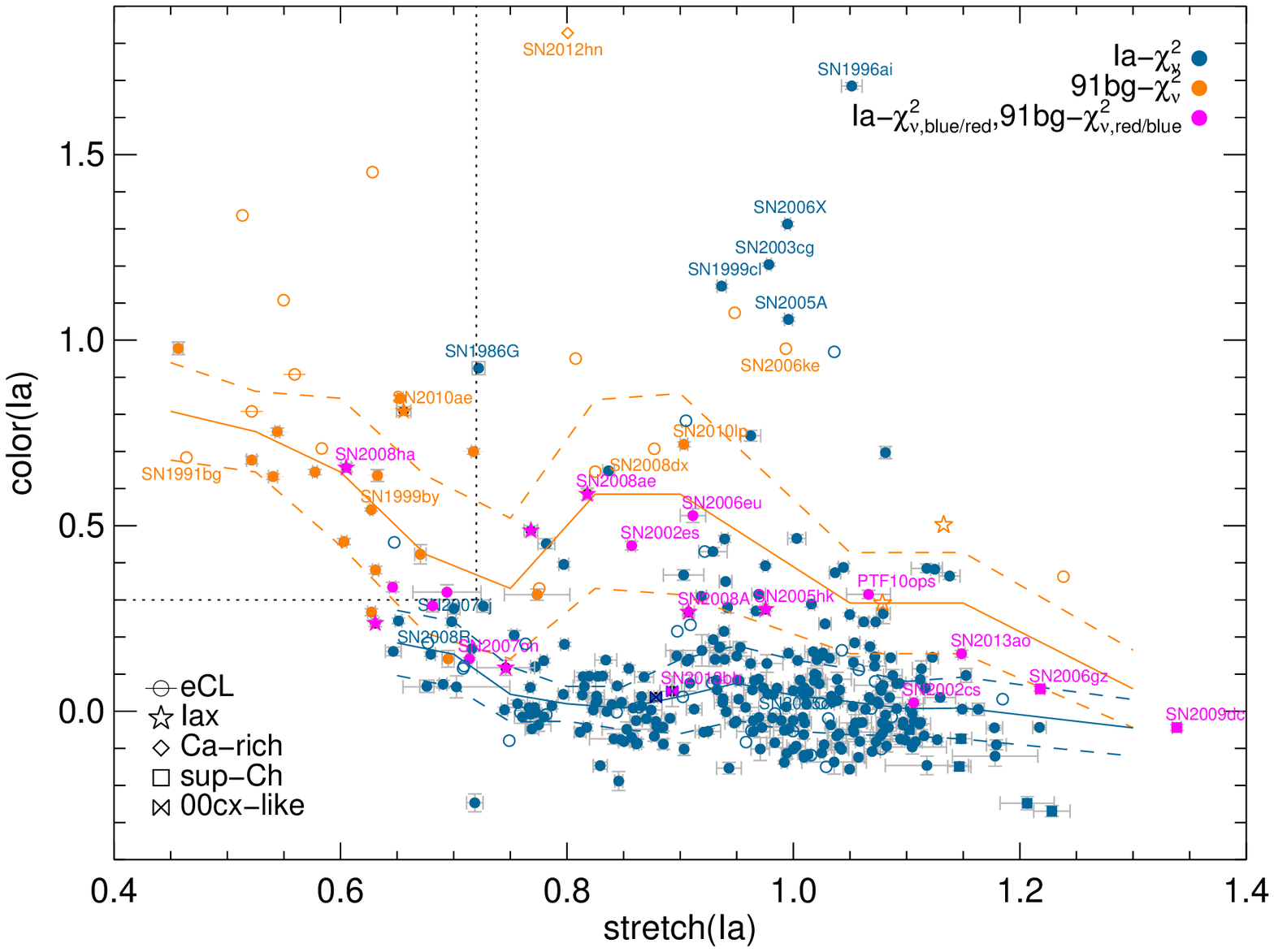}
\includegraphics[width=0.75\linewidth]{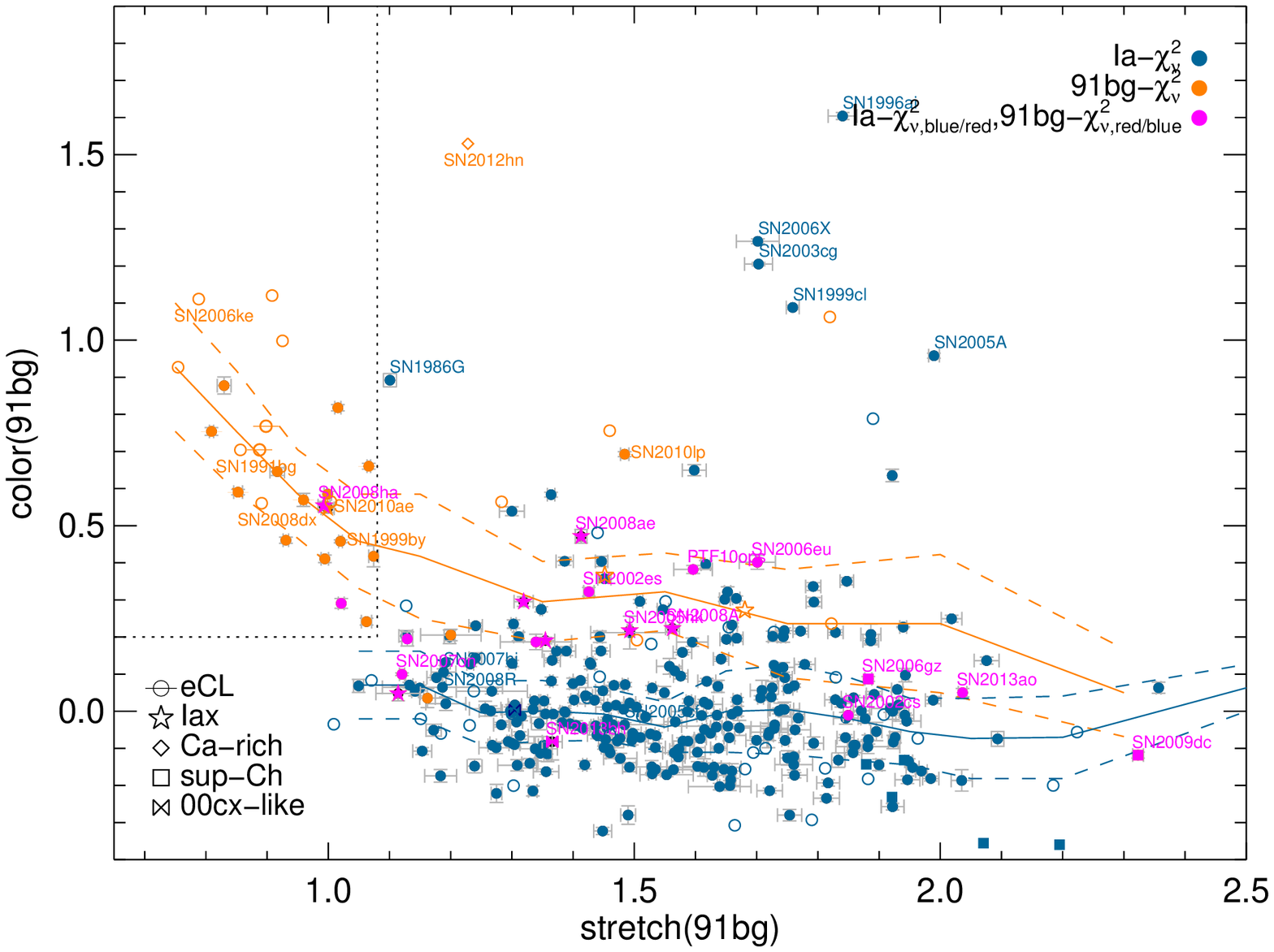}
\caption{a) SiFTO colors vs stretch for the normal template fits (\emph{top}), and b) 91bg template fits (\emph{bottom}). Same symbols and colors as in figure~\ref{dm25max}. The solid lines represent the median of the photometric normal (91bg-like) sample in blue (orange) according only to the overall fits (criterion 1). Dashed lines are the standard deviation on the median for both samples. The vertical and horizontal dotted lines denote the 91bg-like SN~Ia region. }
       \label{colst}
\end{figure*}

We explore in figure~\ref{colst} the range of parameters obtained with our fits comparing SiFTO color $\mathcal{C}$ with stretch $s$ obtained with both templates: normal and 91bg-like. Since SiFTO fits every filter flux scale independently, both colors, measured with normal and with 91bg templates, should in principle be comparable. For stretch, on the other hand, since a $s=1$ has different meanings in both cases, the values will be quite different but the stretches should be quite correlated. We find indeed following relation for stretch: $s_{\mathrm{Ia}}=(1.676\pm0.001)\times s_{\mathrm{91bg}}-(0.082\pm0.002)$ with RMS$=0.16$, and for color: $\mathcal{C}_{\mathrm{Ia}}=(1.007\pm0.003)\times\mathcal{C}_{\mathrm{91bg}}-(0.040\pm0.001)$ with RMS$=0.06$. Which parameters one should use is defined by the fit quality: e.g. if an object has better 91bg template fit, one should use the parameters obtained with the 91bg template fit, and vice versa. 
 
\begin{figure*}[htbp]
\centering
\includegraphics[width=0.75\linewidth]{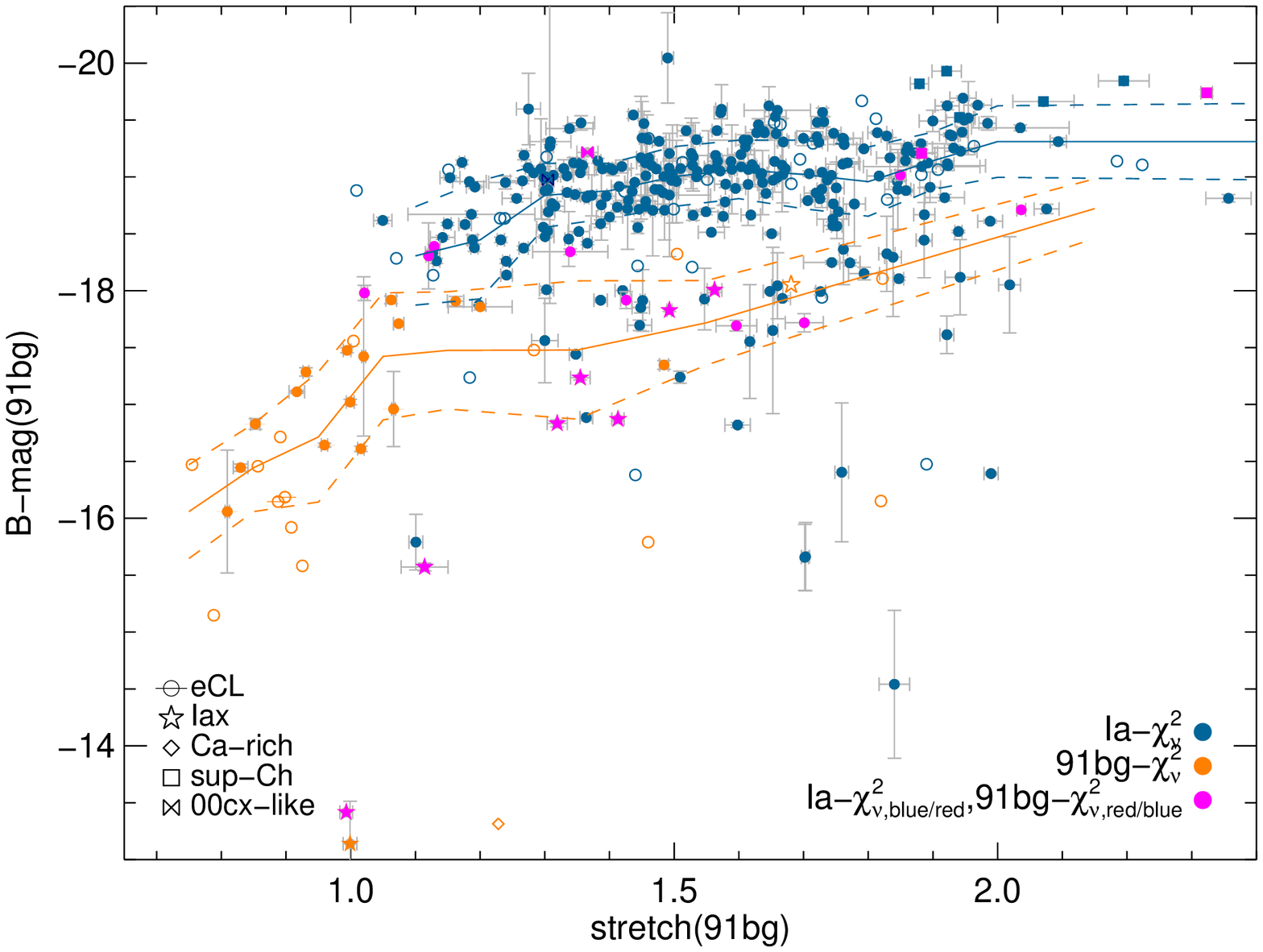}
\includegraphics[width=0.75\linewidth]{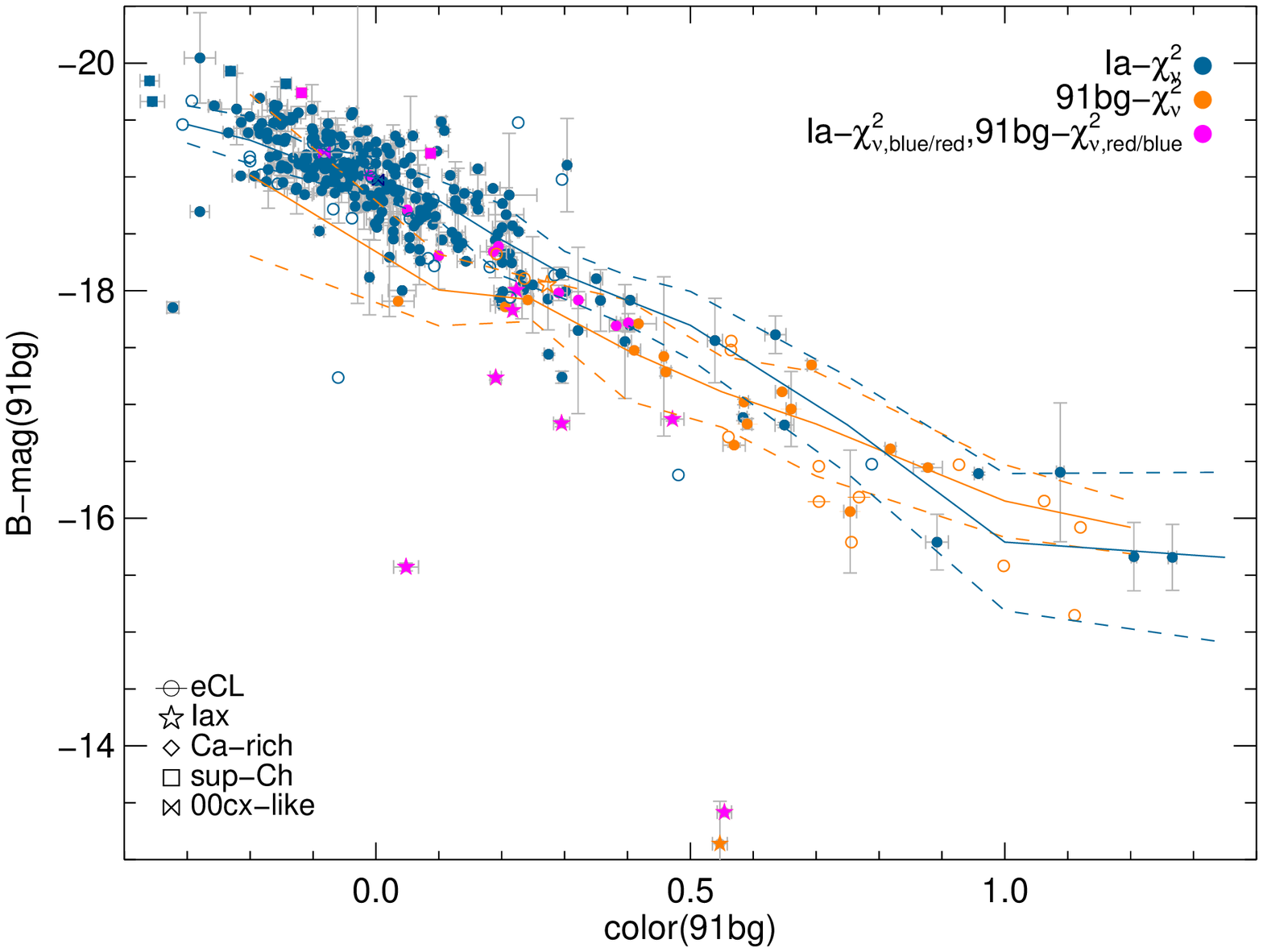}
\caption{a) SiFTO $B$-magnitude vs stretch (\emph{top}), and b) SiFTO $B$-magnitude vs color (\emph{bottom}) for 91bg template fits. Magnitudes are corrected for distance and MW extinction. Same symbols and colors as in figure~\ref{dm25max}. Solid lines show the median of the photometric normal (91bg) sample in blue (orange) according only to the overall template fits (criterion 1). Dashed lines are the respective standard deviation on the mean.}
       \label{magstcol}
\end{figure*}

We also show the magnitude-stretch and magnitude-color obtained with SIFTO from the 91bg template fits in figure~\ref{magstcol}. These magnitudes are only corrected for distance assuming a standard cosmology for SNe in the Hubble flow ($z>0.01$) or different estimates from the NASA/IPAC Extragalactic Database (NED) when closer. We can see that most of SNe~Ia, including normal and 91bg-like, follow a narrow trend of magnitude versus color, regardless of the nature of the color: intrinsic or reddened. Some SNe~Iax are the only exception being very faint for their colors. The different aspects of the SN~Ia sub-groups shown in these figures will be further developed in the next sections.

\subsection{SN~1991bg-like SNe~Ia}\label{91bg}

We now focus on the SN~1991bg-like sample defined as objects with better overall, blue and red 91bg template fits (orange box in figure~\ref{brchisq}), which represent the classical objects similar to SN~1991bg. In both panels of figure~\ref{colst}, we can see that this group spans the shortest stretch range, between $s_{\mathrm{Ia}}\simeq0.4-0.75$ (or $s_{\mathrm{91bg}}\simeq0.7-1.1$). However, we note that a cut based solely on normal stretch, as done in previous studies, would be insufficient to define classical 91bg-like SNe~Ia photometrically. Some SNe~Ia like SN~2008R have a fast evolving light-curve, i.e. low stretch, but are better fit with a normal SN~Ia template. Some normal low-stretch SNe~Ia do exist and are characterized for having bluer colors than normal SNe~Ia. Therefore, a color cut would also be needed to ensure an almost uncontaminated population. Using a color cut of $\mathcal{C}_{\mathrm{Ia}}>0.3$ (or $\mathcal{C}_{\mathrm{91bg}}>0.2$) and a stretch cut of $s_{\mathrm{Ia}}<0.72$ (or $s_{\mathrm{91bg}}<1.08)$ (shown with dashed lines in figure~\ref{colst}) instead of the above $\chi^2$ criteria, we recover most of the real 91bg-like candidates leaving out all other SNe~Ia except for two SNe~Iax of very low stretch. Using exclusively a cut with the 91bg stretch, $s_{\mathrm{91bg}}<1.08$, is a better discriminator that almost does not require the extra color cut needed when using the normal stretch. We note that a potential problem of a simple color and stretch cut instead of our $\chi^2$ fit analysis would be the presence of highly reddened normal SNe~Ia at low stretch mimicking typical 91bg-like SNe~Ia.

It is worth mentioning that the transition region between photometric normal and 91bg-like objects is well populated in both figures ~\ref{brchisq} and ~\ref{colst}a, with no evidence of a clear gap as hinted in \citet{Phillips11}. The few 91bg-like objects found in the defined normal region have very similar $\chi^2_{\nu}$ for both templates. This is the case for SN~2007on for example, a photometric 91bg-like SN~Ia with a very similar fit quality with a normal template. On the other hand, SN~2007hj is a photometrically normal SN~Ia with basically the same fit quality with both templates. In fact, there are in total 8 objects with $|\chi^2_{\nu}(\mathrm{Ia})-\chi^2_{\nu}(\mathrm{91bg})|<1$, of which 5 lie in the transition region, i.e. $0.68<s_{\mathrm{Ia}}<0.80$. Other examples include SN~1986G, a low-stretch ($s_{\mathrm{Ia}}=0.72$), highly reddened from extincion in the host galaxy. This SN has slightly better normal templates fits with indications of shoulder and secondary maxima in the red-bands. This is ultimately confirmed through NIR observations \citep{Phillips11}. These transition objects hint towards a smooth transition from 91bg-like to normal SNe~Ia. Nonetheless, one can arguably see a slight gap in the color-stretch figure for the 91bg template (figure~\ref{colst}b) and the behaviour of the 91bg-like sample for the color-stretch relations, as well as for the relations between magnitude-stretch (see figure~\ref{magstcol}a) present different slopes than for the normal population. We will investigate this issue further with a cluster analysis in section~\ref{sec:cluster}.

\citet{Folatelli13} define an ``extreme cool'' (eCL) spectroscopic sample based on the pseudo-equivalent width of the \ion{Mg}{2} line complex around 4300\AA, which includes \ion{Ti}{2} lines. The more extreme 91bg-like objects have the strongest features and are therefore eCL objects. We identify those objects in our figures and we note that, as expected, they have the largest $\chi^2_{\nu}$ difference between normal and 91bg template fits, giving further evidence of the power of this technique. 

Finally, it is worth noting that two objects, SN~2006ke and SN~2008dx, spectroscopically classified as SN~1991bg-like with SNID in \citet{Silverman12} and that we identify as 91bg-like candidates according to the overall fit (but we do not have sufficient data in blue-bands to confirm them) lie in the proper region defined in the upper left of figure~\ref{colst} for the 91bg fits (\ref{colst}b) but not for the normal fits (\ref{colst}a). This confirms that one should use the parameters obtained with the best template.

\subsection{Slowly evolving SN~1991bg-like SNe~Ia}\label{ptf10ops}
The present study has revealed that many peculiar SNe~Ia bear some light-curve similarities with typical 91bg-like SNe~Ia and that they are classified as such with a simple single criterion. The clearest example of this are the high-stretch 91bg-like SNe~Ia: PTF~10ops \citep{Maguire11} and SN~2010lp \citep{Pignata14,Kromer13}, which have wide light-curves, yet 91bg-like spectra and red colors. This set of objects are classified as 91bg-like according to criterion 1 of our photometric method: they have better overall light-curve fits with a 91bg template. Furthermore, SN~2010lp has both blue- and red-band fits that are better with a 91bg template (criterion 2) as well. For PTF~10ops, there is only one blue band, $g\prime$ filter, with few data points and whose fit is  $\chi^2_{\nu,\mathrm{blue}}(\mathrm{Ia})=0.36$ vs  $\chi^2_{\nu,\mathrm{blue}}(\mathrm{91bg})=0.40$, basically identical, so it is difficult to disentangle between a peculiar SN~Ia such as SN~2002cx or a high-stretch 91bg-like SN~Ia. However, if we take into account the region denoted in purple in quadrant II of figure~\ref{brchisq} as a delimiter of SNe~Iax, and we include the purple dots below as possible 91bg-like objects, we add three more SNe~Ia to the 91bg-like objects: PTF~10ops, SN~2006eu and SN~2007ba. The latter one is a transitional object with $s_{\mathrm{91bg}}=1.13$ ($s_{\mathrm{Ia}}=0.68$) for which SNID gives a 91bg-like classification. SN~2006eu is a higher-stretch object ($s_{\mathrm{Ia}}=0.91$,$s_{\mathrm{91bg}}=1.70$) with a normal SNID classification; the CfA classification\footnote{\url{http://www.cfa.harvard.edu/supernova/RecentSN.html}}, however, signals its similarity to a 91bg-like object. 
Another object, SN~1999bh with $s_{\mathrm{91bg}}=1.46$ and $\mathcal{C}=0.76$, is typed as 91bg-like according to the overall fits but not enough pre-maximum data in the blue is available to study it further. SNID, however, gives 91bg-like matches making this SN another potential member of this group, although it also finds some matches to SN~2006bt (see \S~\ref{06bt}). \citet{Ganeshalingam12} also mention the peculiarity of this object. This would imply that we identify 3-4 high-stretch SNe with photometric 91bg-like characteristics according to criteria 1 and 2. No other similar SNe~Ia are known in the low-$z$ literature, suggesting that they are rare or that they do occur in remote regions \citep{Maguire11}, and therefore missed in historic galaxy targeted surveys. 

\subsection{Type Iax supernovae}\label{iax}
\begin{figure*}[htbp]
\centering
\includegraphics[width=0.49\linewidth]{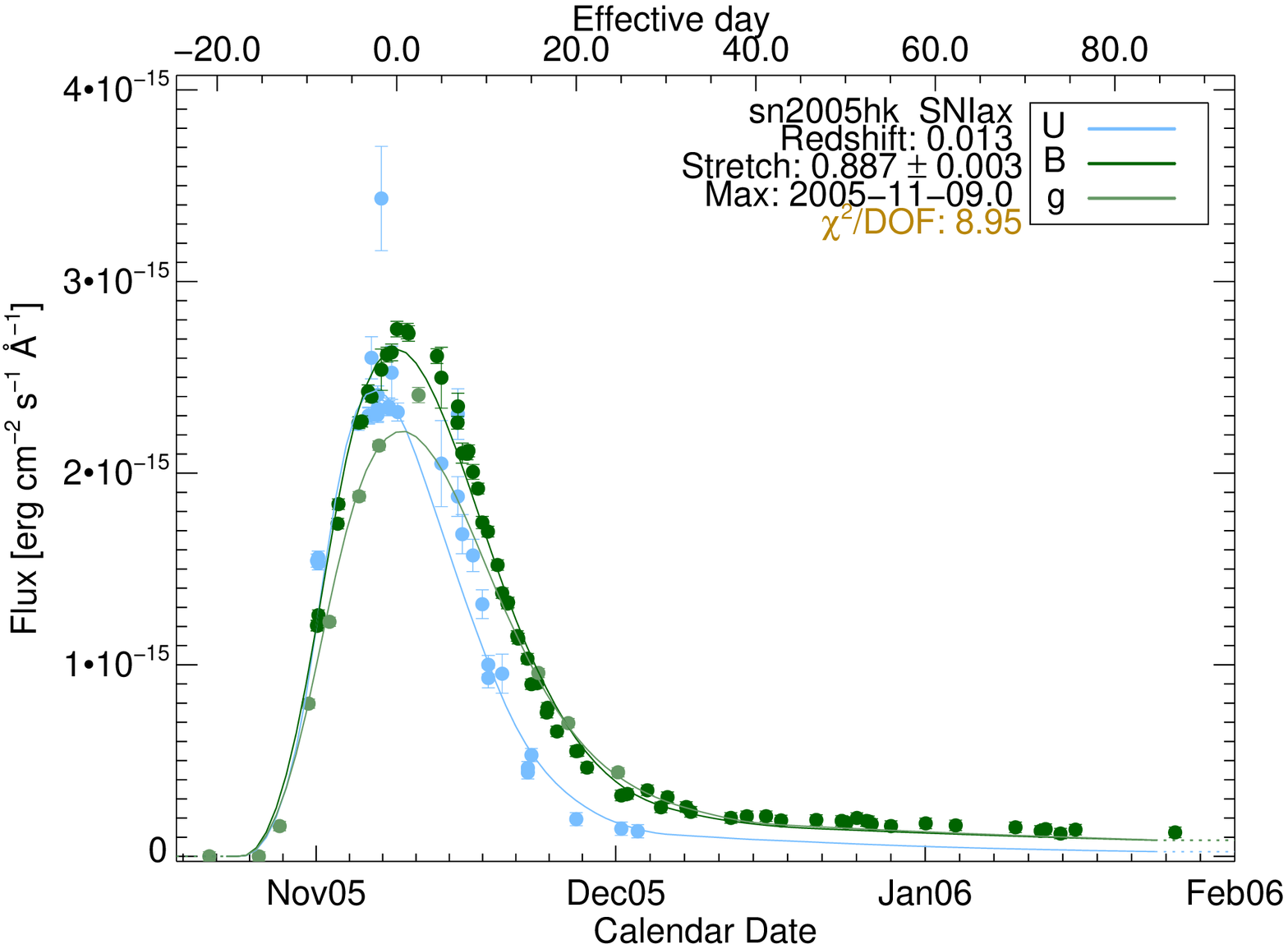}
\includegraphics[width=0.49\linewidth]{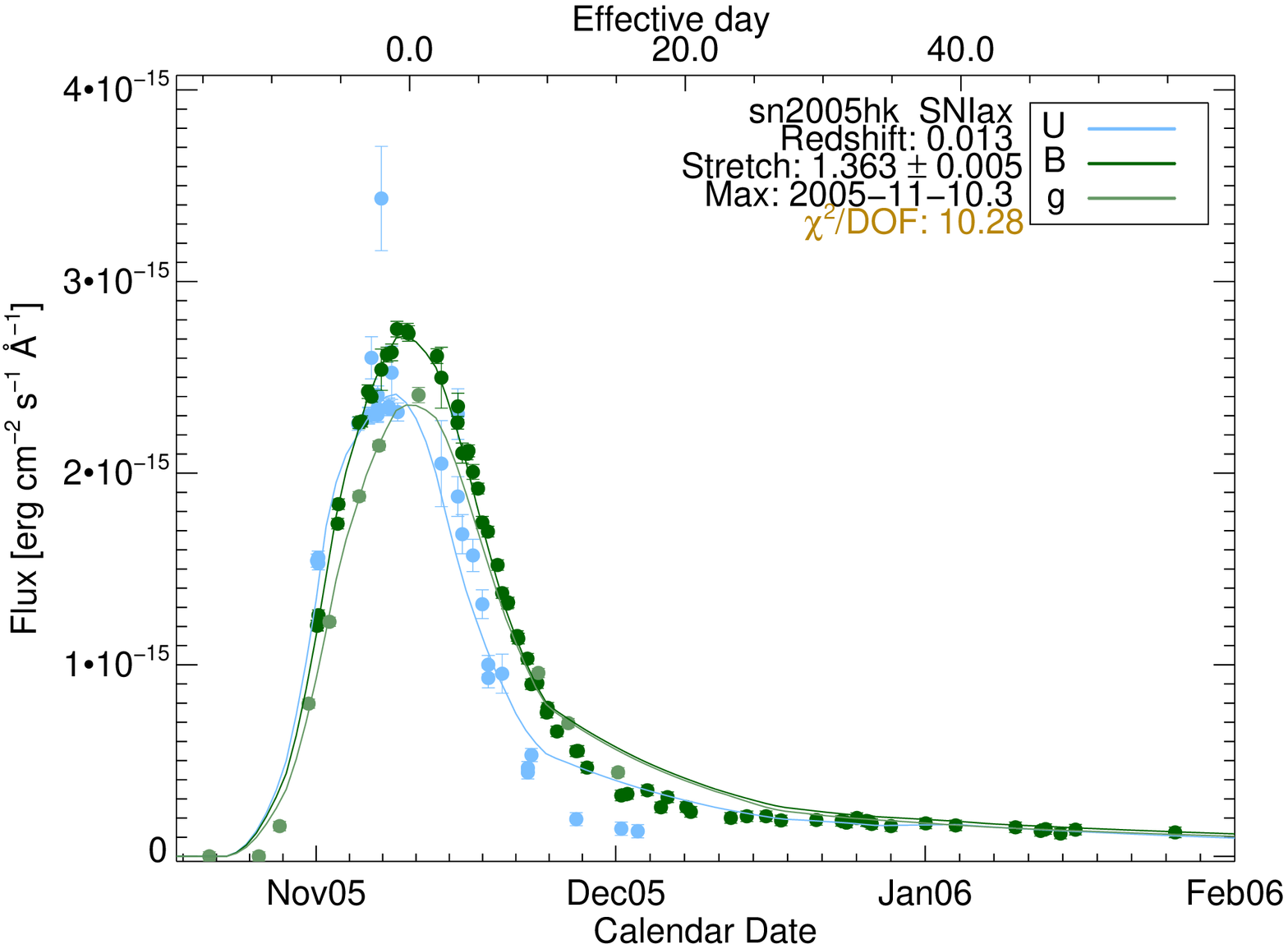}
\includegraphics[width=0.49\linewidth]{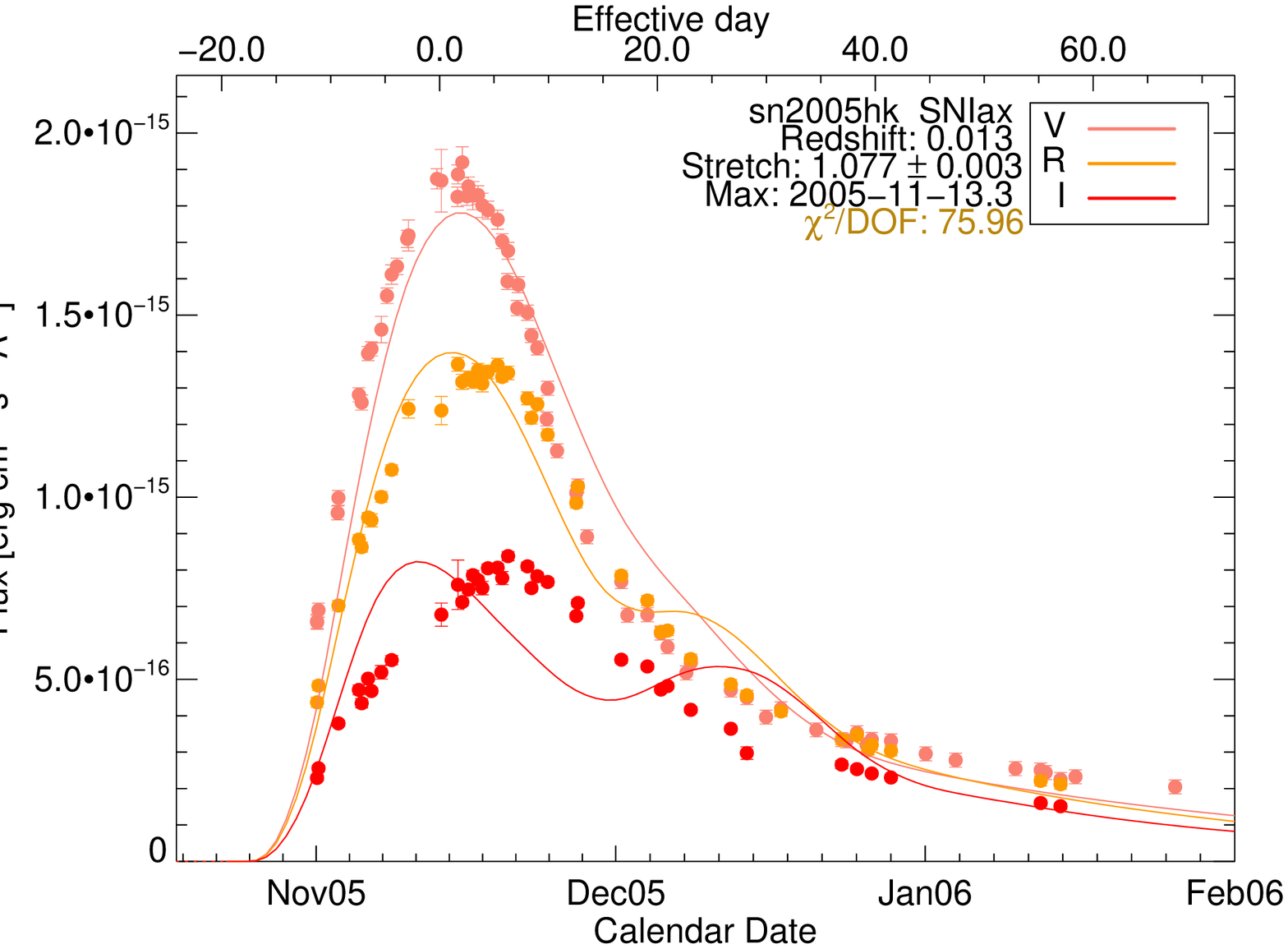}
\includegraphics[width=0.49\linewidth]{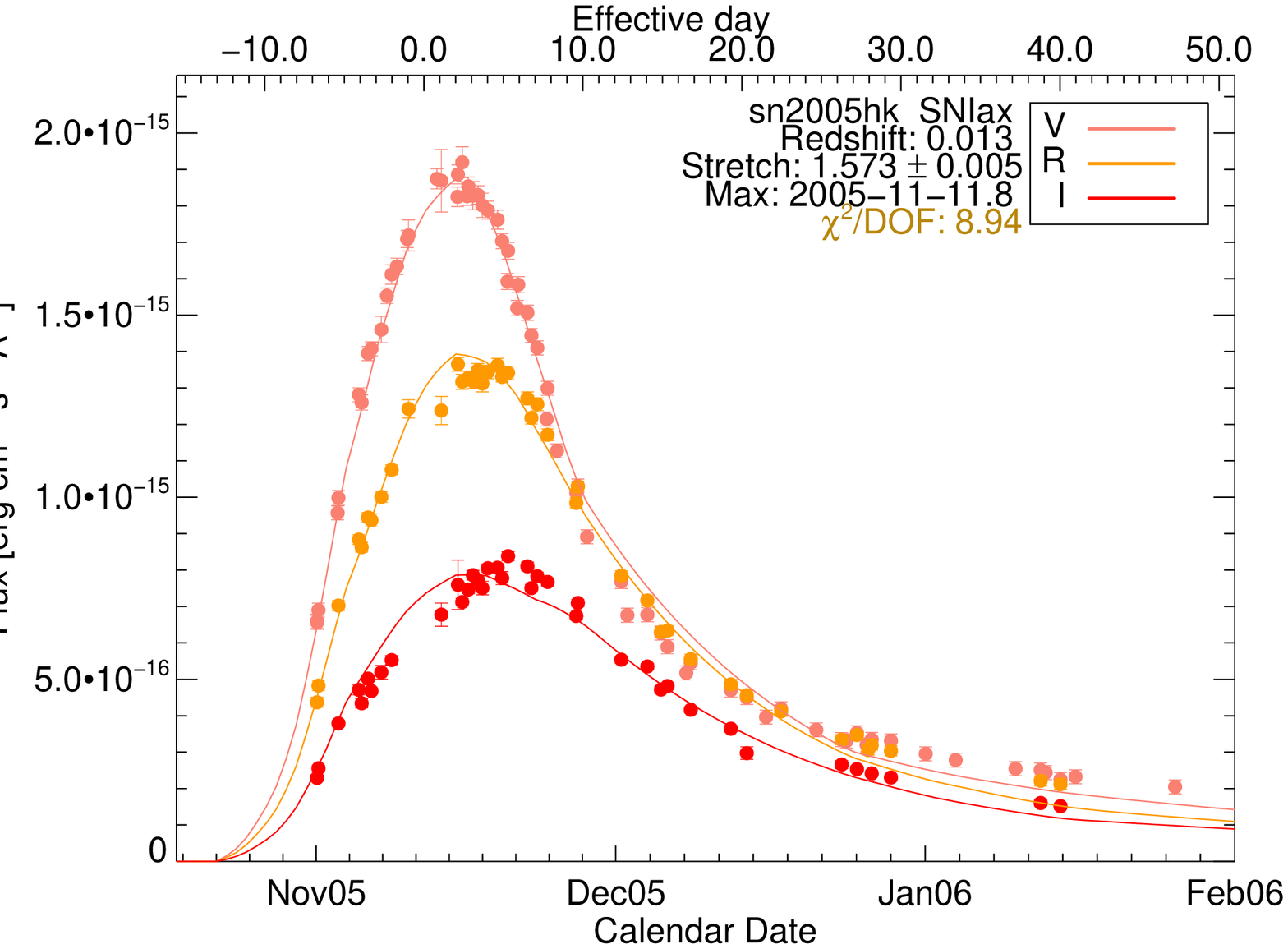}
\caption{SiFTO fits to the light-curve of SN~2005hk with a normal SN~Ia template (\emph{left}) and with a 91bg SN~Ia template (\emph{right}) using exclusively blue-bands (\emph{upper}) and exclusively red-bands (\emph{lower}).}
       \label{iaxplots}
\end{figure*}

An important outcome of the initial overall $\chi^2_{\nu}$ comparison of the fits with the two different templates is the identification of all 10 SNe~Iax in the literature that pass the light-curve cuts. Including then the blue- and red-band $\chi^2_{\nu}$ comparison, we can differentiate between real 91bg-like objects and the remaining 7 SNe~Iax after the new cuts, with the exception of SN~2010ae which has better 91bg template fits in all cases as well as a low stretch. So, from a photometric perspective, SNe~Iax appear as a mixture between normal and 91bg-like SNe~Ia: they have a blue-band behaviour that is more similar to standard SNe~Ia but have red-band light-curves much more similar to 91bg-like SNe~Ia. In figure~\ref{iaxplots}, we show SN~2005hk as an example of all four fits: blue-band and red-band fits for normal and 91bg templates. One can see that although the blue-band fit agrees more with a normal template, the fit is far from perfect. This behaviour is found in other SNe~Iax, even when only one blue filter is available, like for SN~2002cx. Even with the use of templates that weren't designed for SNe~Iax, we are able to identify them quite well.

SN~2010ae is an extreme of the SN~Iax population \citep{Stritzinger14} with very low stretch but also red colors. The difference in $\chi^2_{\nu}$ for both templates is quite large, also for the blue-band fits. This is a striking result given that SN~2008ha, another extreme case of the population, does fall into the SN~Iax category through the blue-band fits. Both however are very faint (even taking into account extinction, see \citealt{Stritzinger14}), far off the classical 91bg-like population, as shown in figures~\ref{magstcol}, so they can easily be identified. Two more SNe~Iax, SN~2011ay and SN~2012Z, are photometrically identified as 91bg-like candidates with criterion 1 but not enough data in the blue bands permitted to evaluate criterion 2.

On the other hand, we do find another candidate that is not classified as SN~Iax in the literature: SN~2002es. Classifying it with SNID, we find that it is a 91bg-like object with $s_{\mathrm{Ia}}=0.86$ and $s_{\mathrm{91bg}}=1.43$; no match to any SN~Iax is found. \citet{Ganeshalingam12} note the peculiarity of this object and discuss the similarities to 91bg-like and to 02cx-like objects. Therefore, it seems this object might be in between SNe~Iax and slowly evolving 91bg-like objects, such as PTF~10ops. Alternatively, it could make part of another similarly peculiar group, SN~2006bt-like objects, that will be discussed later. 


\subsection{Super-Chandrasekhar SNe~Ia}
Besides SNe~Iax, another 8 SNe~Ia have better overall and worse blue- or red-band 91bg template fits. Two of those catch one's attention: SN~2006gz and SN~2009dc. These two supernovae are super-Chandrasekhar SN~Ia candidates \citep[e.g.][]{Hicken07,Maeda09,Taubenberger11,Hachinger12,Kamiya12}. Taking a closer look at their light-curves, one can understand why this happens: the red-band part is better fit with the 91bg template. Since super-Chandrasekhar SNe~Ia have characteristic $I/i$-band light-curves where the secondary maximum happens earlier than for normal SNe~Ia, the two maxima merge and create almost a plateau with an unique elongated maximum. SN~1991bg-like SNe~Ia have also only one maximum, that, even of shorter duration, can approximate super-Chandrasekhar light-curves better. The same occurs in the $R/r$ band where the shoulder of super-Chandrasekhar almost disappears, simulating the behaviour of 91bg-like objects (see figure~\ref{supplots}). Investigating the classical super-Chandrasekhar SN~2003fg \citep{Howell06}, which was not originally included in our analysis due to its higher redshift ($z=0.24$), we find that it is also better fit with a 91bg template for the overall but also the blue- and red-band light-curves. However, given its high stretch ($s_{\mathrm{91bg}}=1.99$, $s_{\mathrm{Ia}}=1.30$), one can easily tell it apart from regular 91bg-like and PTF~10ops-like objects. SN~2007if \citep{Scalzo10}, on the other hand, does not have pre-maximum data to perform our analysis. In addition to these supernovae, \citet{Scalzo12} present 5 super-Chandrasekhar candidates. Our technique does not pick these up at all. This is due to their wider definition more similar to typical SN~1991T-like objects. One can indeed see that these light-curves have clearly distinct secondary maxima in $I/i$ and definite shoulders in $R/r$. Our method suggests that only the most extreme cases, the standard super-Chandrasekhar SNe~Ia, have photometric similarities to other sub-classes as 91bg-like and SN~Iax objects. 

This method seems then capable of distinguishing also super-Chandrasekhar objects. How can one separate them from SNe~Iax since they are all identified in the same manner? When one looks at the stretch of these objects, they are beyond $s_{\mathrm{91bg}}>1.8$ ($s_{\mathrm{Ia}}>1.2$). Some super-Chandrasekhar candidates identified by \citet{Scalzo12} have such wide light-curves but interestingly, their colors are bluer ($\mathcal{C}<-0.12$) as opposed to SN~2006gz ($\mathcal{C}=0.06$) and SN~2009dc ($\mathcal{C}=-0.05$). SNe~Iax, on the contrary, all have $s_{\mathrm{91bg}}<1.6$ ($s_{\mathrm{Ia}}<1.05$). As is the case for 91bg-like objects (\S~\ref{91bg}), this division is clearest with the 91bg instead of the normal stretch.

Another interesting object, SN~2002cs, is identified with our method in a similar way to the super-Chandrasekhar candidates. It also has high stretch, $s_{\mathrm{91bg}}=2.04$ ($s_{\mathrm{Ia}}=1.15$), red color ($\mathcal{C}=0.16)$ and it shows small shoulders instead of a secondary maximum in the $i$-band light-curve and no $r$-band shoulder at all. It is better fit in the red with a 91bg template but not in the blue. However it is not as bright as typical super-Chandra (mag$_B=-19.05$). An additional tentative super-Chandra candidate, SN~2009li, at $s_{\mathrm{91bg}}=1.82$ ($s_{\mathrm{Ia}}=1.24$), $\mathcal{C}=-0.15$ and mag$_{B}=19.93$, did not have enough data to probe only the blue-bands.

\subsection{SN~2006bt-like objects}\label{06bt}
\begin{figure}[htbp]
\centering
\includegraphics[width=1.0\linewidth]{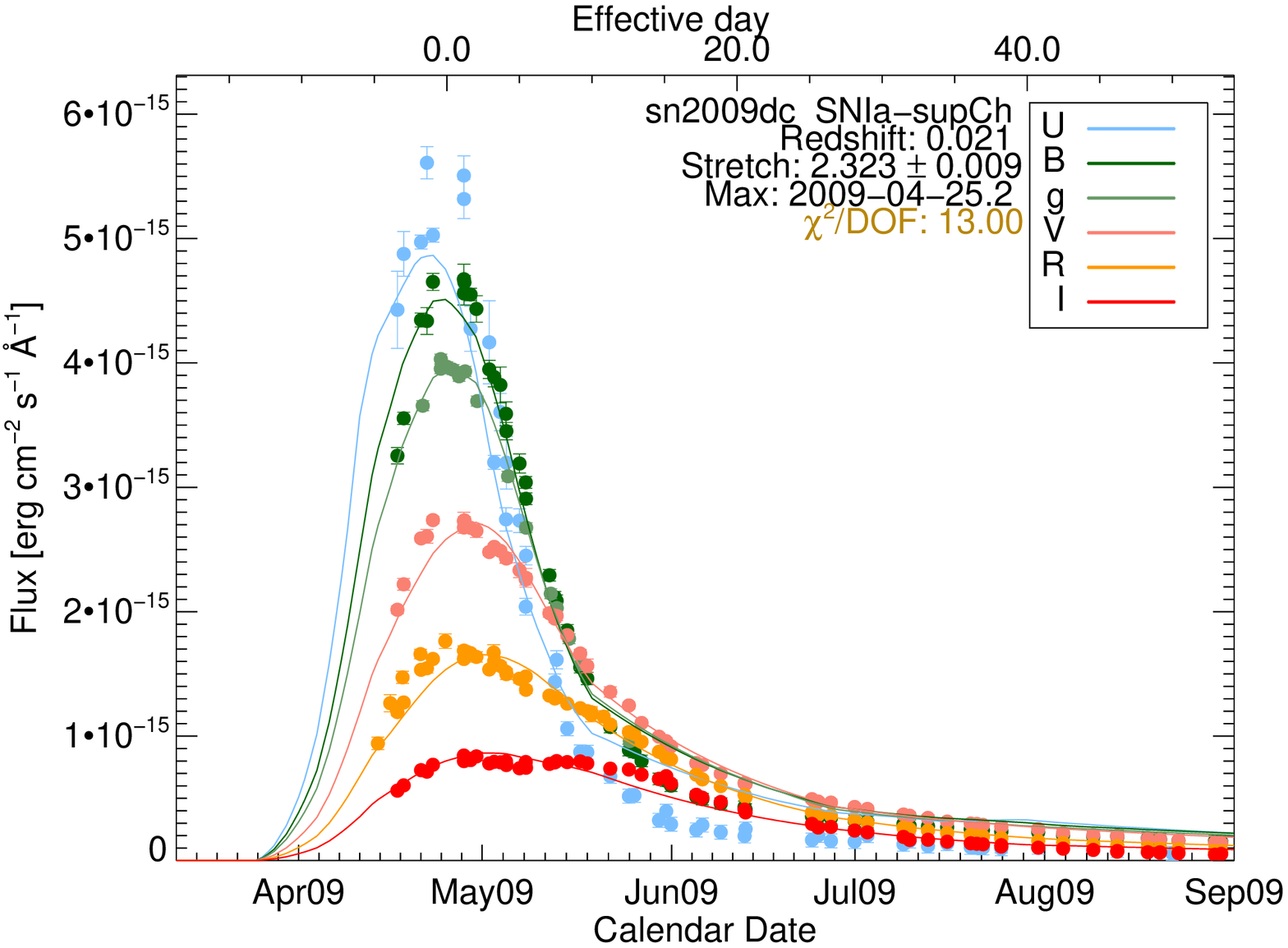}
\includegraphics[width=1.0\linewidth]{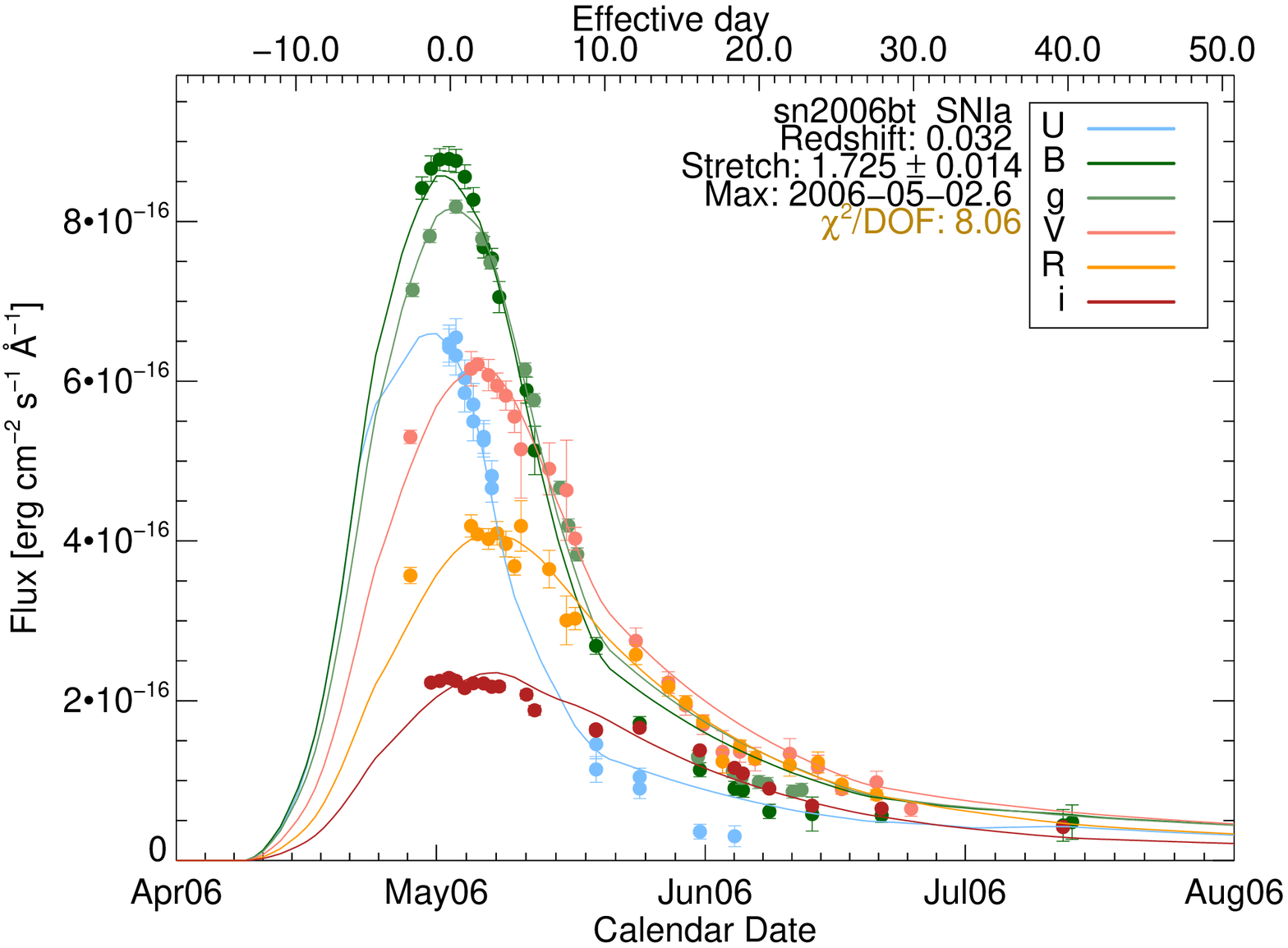}
\includegraphics[width=1.0\linewidth]{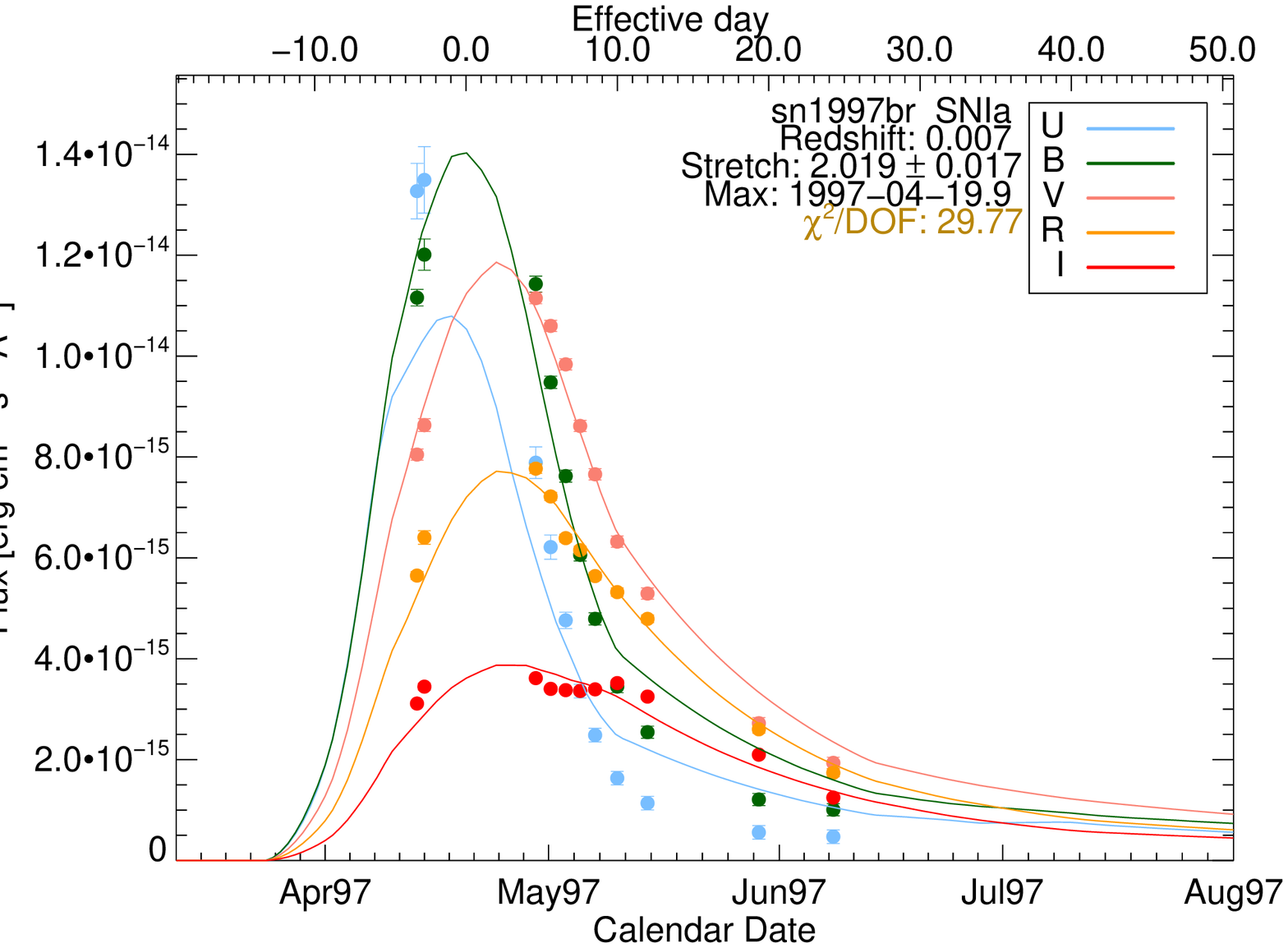}
\caption{SiFTO fits to the light-curves of SN~2009dc, SN~2006bt and SN~1997br with a 91bg SN~Ia template.}
       \label{supplots}
\end{figure}

Some of the objects presented in last sections, like the latter super-Chandrasekhar candidate, SN~2002cs, are reminiscent of the peculiar SN~Ia presented by \citet{Foley10}, SN~2006bt. SN~2006bt has spectra similar to 91bg-like SNe~Ia, red colors,  no prominent secondary maxima and it occured in an early-type galaxy, yet it has a slowly declining light-curve. Taking a look at the fits of this SN, we see that although the overall fit is better with a normal template, the red-band part is better fit with a 91bg template. In fact, there are several objects for which the overall fit is better with a normal template but the red-band part is consistent with the 91bg template. All these objects are shown in blue in quadrant II of figure~\ref{brchisq}, as opposed to most SNe~Iax and super-Chandrasekhar that, with better overall 91bg templates fit, are also in that same quadrant but in purple. Investigating these 9 objects, we find that 4 of them (SN~1986G, SN~2002dl, SN~2007fr and SN~2007hj), with $0.68<s_{\mathrm{Ia}}<0.72$ ($1.05<s_{\mathrm{91bg}}<1.20$), are transitional objects between classical 91bg-like and normal SNe~Ia. In fact, their spectra do have some SNID matches to 91bg-like objects as does the CfA classification. On the other hand, the other 5 objects (SN~1981D, SN~1989A, SN~1997br, SN~2001bf and SN~2006bt) have higher stretch $0.9<s_{\mathrm{Ia}}<1.1$ ($1.4<s_{\mathrm{91bg}}<2.0$) and red colors ($\mathcal{C}>0$). Of these SNe, the last three have sufficient coverage in $R/r$ and/or $I/i$ to inspect the behaviour more closely: SN~2001bf has a secondary maximum too close to the first maximum in the $I$-band, SN~1997br shows the maximum much earlier almost forming a single elongated plateau with the first maximum, whereas for SN~2006bt this plateau is almost gone (see figure~\ref{supplots}). 

This means that, from the light-curve behaviour perspective, super-Chandrasekhar SNe~Ia and these 06bt-like SNe~Ia may form a same family going from the ones where the two maxima join in a long plateau in red bands, including typical super-Chandrasekhar SN~2006gz and SN~2009dc, but also SN~2002cs, then there is a transition of steeper and shorter plateau durations with SN~2006bt, to finish-up with SNe~Ia that show a weak or very early secondary maximum like SN~1997br and SN~2001bf. These two last ones are more similar to normal SNe~Ia and could be a link between super-Chandrasekhar and normal SNe~Ia. In particular SN~2001bf is not as red as all the previous objects. Finally, it is interesting noting that SNID gives a normal Ia classification to SN~2001bf, a 91T-like classification to SN~1997br, and some matches to SN~2006bt are consistent with SN~1986G. This validates that the wide variety of objects that we find to be similar photometrically are partly similar spectroscopically as well. 

Finally, the possible high-stretch 91bg-like candidate found in \S~\ref{iax}, SN~2002es, has several interesting spectrocopic matches to SN~2009dc according to GELATO, whereas the one found in \S~\ref{ptf10ops}, SN~1999bh, has SNID matches to SN~2006bt, making them other potential members of the 06bt-like instead of the PTF~10ops-like group. And some additional objects without enough data to do blue-band fits, SN~2003ae and SN~2007kd, are also typed as a photometric 91bg-like SN~Ia according to criterion 1, while spectroscopic classificators indicate it to be a normal SN~Ia. These ones could be like SN~2001bf.

\subsection{Other peculiar SNe~Ia}

\emph{SN~2000cx-like SNe~Ia:}
SN~2000cx-like objects are another class of peculiar, very rare SN~Ia objects with only two reported members in the literature, SN~2000cx and SN~2013bh \citep{Silverman13}. We obtain very different results with our technique for both. SN~2000cx is quite compatible with a normal SN~Ia template whereas SN~2013bh, having bad fits in both, is slightly more consistent with a 91bg-like SN~Ia for the overall and blue-band fits but not for the red-band fits (one of the few objects in quadrant IV in figure~\ref{brchisq}). This SN has a blue color, lying far off from SNe~Iax, 06bt-like objects and super-Chandrasekhar SNe~Ia in the color-stretch diagram (figures~\ref{colst}). With such small statistics, it is hard to conclude much about them and their possible identification, however if such a blue object, $\mathcal{C}=-0.08$, at mid-stretch, $s_{\mathrm{91bg}}=1.37$, passing criterion 1 but not 2, is found, it is probably a peculiar SN~Ia that doesn't fit any of the other categories, i.e. it is  a SN~2000cx-like object.


\emph{Ca-rich transients:}
Ca-rich transients are an emerging class of objects, possibly thermonuclear in origin \citep{Perets10,Kasliwal12}. Only one of five, SN~2012hn, passes the light-curve cuts and has proper SiFTO fits. It is typed as a 91bg-like candidate with the overall fit but not enough pre-maximum data in the blue allows a further identification. If we were to force the fit anyway, we would obtain a better normal blue-band fit. Its extreme red color ($\mathcal{C}=1.53$) is beyond the extreme cool 91bg-like SNe~Ia and the most reddened SNe~Ia in the top of figure~\ref{colst}. Additionally, if one relaxes the light-curve coverage cuts, one additional SN, PTF~09dav, a peculiar SN~Ia with 91bg-like characteristics \citep{Sullivan11a}, has proper fits and is also typed as a 91bg-like SN~Ia with also very red color ($\mathcal{C}=0.63$). It is then arguably possible to also include Ca-rich transients into the wide variety of peculiar SNe identified with the method presented here. An intriguing transient, SN~2006ha, has similar characteristics: it passes criterion 1 but it does not pass the cuts to check criterion 2, it has a very red color ($\mathcal{C}=1.06$) and SNID and GELATO tend to find core-collapse or AGN matches.


\section{Discussion}\label{discussion}
\subsection{Core-collapse contamination}\label{cont}
The versatibility of the photometric identification method we present may worry the reader as overly capable of fitting many different objects. What if CC~SNe are also incorrectly typed as peculiar SNe~Ia? We investigate this effect in this section by performing the same fits with SiFTO to a large sample of literature CC~SNe (table~\ref{CCphot}) and an additional sample of SNe~II from \citet{Anderson14}. Of the 321 CC~SNe in our sample, we obtain proper fits for 64 of them. Of these, only 11 have better overall normal template fits and 53 have better overall 91bg template fits. This is expected as 91bg-like SNe~Ia are redder and more easily mistaken with CC~SNe, notably SNe~Ibc. 

\begin{figure*}
\centering
\includegraphics[width=0.49\linewidth]{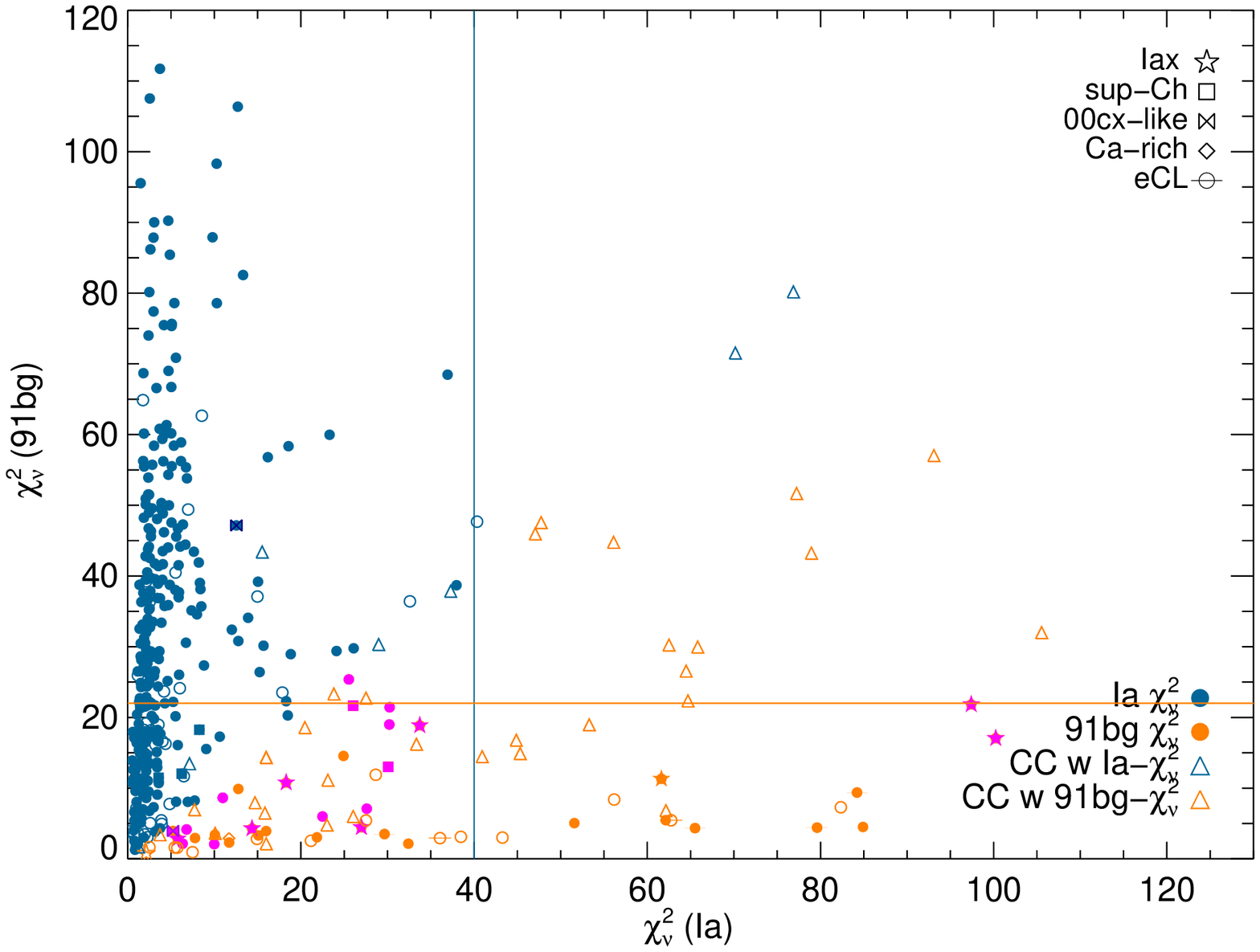}
\includegraphics[width=0.49\linewidth]{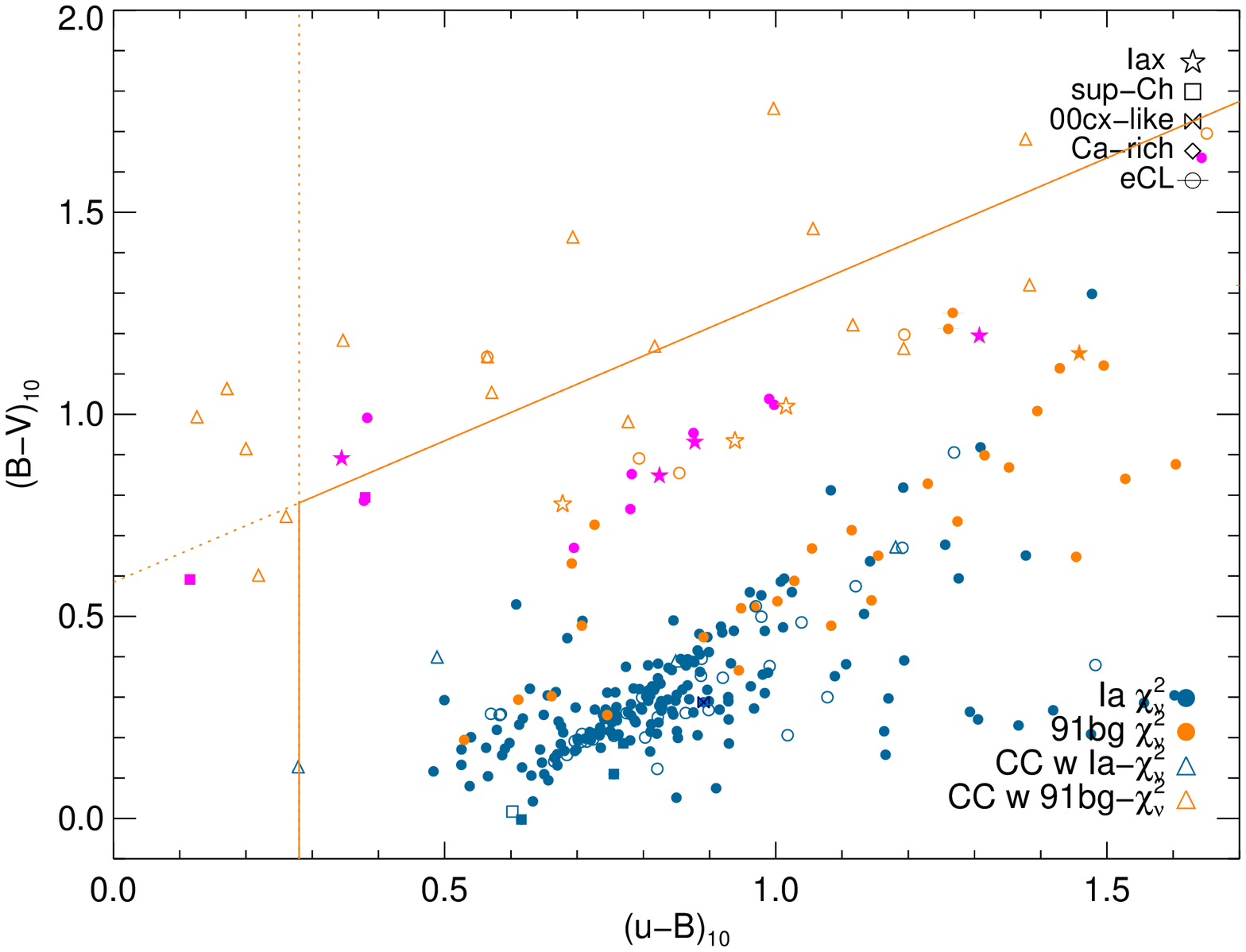}
\caption{Comparison of the fit quality for the normal template versus the fit quality for the 91bg template (\emph{left}) and $B-V$ vs $u-B$ color diagram at 10 days past maximum (\emph{right}) for all SNe~Ia (same symbols as figure~\ref{dm25max}) and for CC~SNe shown as blue and orange triangles when better normal and 91bg template fits were achieved.  Dividing lines serve as selection criteria against CC contamination.}
       \label{chisq1}
\end{figure*}

To deal with CC contamination, we perform several cuts based on the fit quality in the different filter sets, as well as on the magnitude-color relation shown in figure~\ref{magstcol}. In every case, we use a methodology similar to the one presented in section~\ref{bluered} based on the FoM(Ia) and FoM(pec/91bg) to select the best cuts that optimize the number of normal and peculiar SNe~Ia correctly tagged and minimize the number of CC~SNe falsely tagged. First, for those objects, Ia and CC, having better overall normal template fits, we explore cuts in the fit quality in the range $15<\chi^2_{\nu}(\mathrm{Ia})<65$ finding a maximum FoM(Ia) at $\chi^2_{\nu}(\mathrm{Ia})=40$ (see blue line in figure~\ref{chisq1}). This cut eliminates 4 CC~SNe and no SN~Ia. Similarly for objects with better overall 91bg template fits, we search for an optimum cut finding $\chi^2_{\nu}(\mathrm{91bg})=22$ (orange line in figure~\ref{chisq1}). This eliminates 17 CC~SNe and 3 SNe~Ia.

In principle, we could do an equivalent analysis for the fit qualities of blue- and red-band fits searching for cuts that maximize the FoM, but we find that the final FoM does not improve considerably respect to the overall template fits. On the other hand, we can use the fact that many CC~SNe lie outside the typical magnitude relations in figure~\ref{magstcol}. But most effectively, as previously investigated by several authors \citep[e.~g][]{Poznanski02,Perets11}, the evolution of color and the use of color-color diagrams can be a powerful discriminator between SN groups. Since SiFTO does not enforce any color law for SNe~Ia in the fit and allows indepentent flux factors for each band, we study the evolution of $u'-B$, $B-V$, $V-r'$ and $r'-i'$ colors obtained with the best SiFTO fit. We find that a good discriminator is $B-V$ vs $u'-B$ at 10 days after $B$-band maximum. As shown in the right of figure~\ref{chisq1}, CC~SNe are redder in $B-V$ and bluer in $u'-B$ than peculiar and 91bg-like SN~Ia candidates and more so than normal SNe~Ia. Requiring SN~Ia candidates to lie underneath the solid line rejects 26 CC-SNe with only two peculiar SN~Ia discarded. These boundaries are calculated similarly as other cuts: we vary the vertical and diagonal lines until we find a maximum FoM(Ia). Our final sample contains 1 CC contaminant that is typed as a normal SN~Ia according to the overall fit and 4 that are tagged as 91bg-like SNe~Ia according to the overall fit. All of these are stripped-envelope SNe as expected.


We obtain a final efficiency of $\epsilon(\mathrm{Ia})\simeq80\%$ for normal SNe~Ia and a purity of $P(\mathrm{Ia})\simeq99\%$ from CC~SNe and other peculiar SNe~Ia resulting in a FoM(Ia)$\simeq80\%$. This is comparable to other photometric typing techniques as will  be shown in section~\ref{comptype}. For 91bg-like SNe~Ia/peculiar SNe~Ia we obtain $\epsilon(\mathrm{pec/91bg})\sim 65\%$  and a purity of $P(\mathrm{pec/91bg})\sim 85\%$ from CC~SNe and normal SNe~Ia resulting in a FoM(pec/91bg)$\sim 55\%$. These numbers will be discussed in next section, where we investigate how strongly the chosen boundaries and cuts depend on the training sample and how they affect the FoMs.

\subsection{Bootstrap analisys}\label{bootstrap}

The method we use and the boundaries we define are very sensitive to the sample we use to train the technique. In order to test the importance of this, we perform a bootstrap analysis \citep{Efron82} in which different random samples are drawn from the original population to re-do the analysis and calculate the best FoM regions. The bootstrap is done for all SNe~Ia and all CC~SNe, where objects may be absent in a given iteration and others can also be repeated. For every one of the 200 realizations we re-calculate the best cuts presented in sections ~\ref{bluered} and \ref{cont} by finding the maximum FoM box to divide better blue/red-band normal and 91bg template fits (orange box in figure~\ref{brchisq}). We can also find the best $\chi^2_{\nu}$ cuts for the overall normal and 91bg template fits to select SNe~Ia against CC~SNe (lines in figure~\ref{chisq1}) for each iteration, as well as the best factor multiplying the magnitude-color median standard deviation to reject CC contamination (dashed lines in figure~\ref{magstcol}). The final cuts and errors are calculated taking the mean and standard deviation of all realizations and are summarized in table~\ref{fomtable}. The FoMs for each sample with these respective cuts are also presented and the error on the FoM is the FoM calculated at the respective sigma cuts. We emphasize that the quoted FoMs are not the maximum ones, only the ones based on average cuts. Contamination is taken from CC~SNe but also from any other SN~Ia sub-sample.

We obtain a robust classification for normal SNe~Ia with very low contamination in all realizations of the bootstrap analysis. The peculiar SN~Ia identification is more dependent on the sample used for the training and in all cases contains at least 7 stripped-envelope CC~SNe that lower the purity. Interestingly, these are all classified as 91bg-like objects based on overall, blue- and red-band fits. So, based on the fit quality in different bands, CC~SNe are more easily mistaken with 91bg-like than SNe~Iax. However, in the magnitude-color relation, SNe~Iax can have extremely low magnitudes for their colors, emulating CC~SNe. This means that if no magnitude-color relation for objects with overall 91bg template fits is used, the technique is better at classifying SNe~Iax whereas if one enforces such a cut, then 91bg-like objects will be weighted higher. We note that in the case of a pure SN~Ia sample, the blue/red-band cut of section~\ref{bluered} is sufficient to differentiate between the two: 91bg-like and SNe~Iax (and super-Chandra with distinct high stretches). 

\subsection{Outlook for high-redshift typing}

The technique presented can in principle also be used at higher redshifts. SiFTO has been shown to be a robust light-curve fitter across a wide range of redsfhits \citep[e.g.][]{Guy10,Conley11} that works directly with the SED templates in the observer frame, provided an input redshift. For large surveys this redshift can come from template fitting to multiband photometry from the host galaxy \citep[e.g.][]{Ilbert09}, as well as from direct SN light-curve fits \citep[e.g.][]{Palanque-Delabrouille10}. \citet{Sullivan06a} and \cite{Gonzalez11} used a light-curve fitter similar to SiFTO with an additional redshift parameter, that also allows the use of multiple templates such as the 91bg we use here. The particular ``blue'' and ``red'' filter sets will need to be adjusted due to redshift effects. For example, while at $z=0$, a ``blue''filter set consists of filters u$\prime$g$\prime$ and a red filter set of r$\prime$i$\prime$z$\prime$, at $z=0.5$ these will be u$\prime$g$\prime$r$\prime$ and i$\prime$z$\prime$ respectively, while at $z=0.75$ they would move to u$\prime$g$\prime$r$\prime$i$\prime$ and $z\prime$, respectively. Additionally, the cadence and signal-to-noise of the data will affect the quality of the light-curve fit and could induce mistakes in the typing as well. These effects should be considered when applying this technique to high-$z$ surveys; and they will be investigated in a future work.

\begin{table*}[htbp]
 \centering
\caption{Final cuts and errors based on the bootstrap analysis with the respective number of objects of each group passing each cut. Final efficiencies, purities and FoMs for various of our samples are presented.}
 \label{fomtable}
 \begin{threeparttable}
 \renewcommand{\arraystretch}{1.5}
  \begin{tabular}{||>{\centering}m{3.4cm}||c|c|c|c|c|c||}
    \hline
    \noalign{\smallskip}\hline\noalign{\smallskip}
CUT & NORMAL SNe~Ia & PECULIAR SNe~Ia\tnote{$\diamond$}  & 91BG-LIKE SNe~Ia & SNe~Iax &  SUPER-CH SNe~Ia\tnote{$\ast$} & CC~SNe\\
    \noalign{\smallskip}\hline\noalign{\smallskip}
     \hline
None & 602 & 59 & 35 & 13 & 5 & 321 \\
\hline
SiFTO fit \& LC coverage \& $\Delta\chi^2_{\nu}>0$ & 322 & 48 & 29 & 10 & 4 & 64 \\
\hline
$\chi^2_{\nu}(\mathrm{Ia})<(34.0\pm6.65)$    $\chi^2_{\nu}(\mathrm{91bg})<(21.69\pm4.00)$\tnote{\dag} & 304 & 45 & 29 & 8 & 4 & 43 \\
\hline
Color-color relation $(u-B)_{10}>(0.18\pm0.13)$ $(B-V)_{10}<0.7(u-B)_{10}+(0.77\pm0.23)$   & 299 & 44 & 29 & 7 & 2 & 17 \\
\hline
$\Delta\chi^2_{\nu,\mathrm{blue}}>(-0.77\pm2.21)$  $\Delta\chi^2_{\nu,\mathrm{red}}>(2.83\pm3.11)$  & 263  & 31 & 19 & 6 & 2 & 5 \\
\hline
\noalign{\smallskip}\hline\noalign{\smallskip}
\noalign{\smallskip}\hline\noalign{\smallskip}
Efficiency & $81.7\pm0.9$ & $64.6^{+2.1}_{-8.4}$ & $65.5^{+0.0}_{-0.0}$ & $60.0^{+10.0}_{-10.0}$ & $50.0^{+25.0}_{-0.0}$ & - \\
Purity & $99.6\pm0.0$ & $88.6^{+9.8}_{-3.6}$ & $86.4^{+15.9}_{-3.9}$ & $85.7^{+12.5}_{-33.7}$ & $100.0^{+0.0}_{-32.2}$ & -\\
FoM & $81.4\pm0.9$ & $57.2^{+0.0}_{-7.6}$ & $56.6^{+10.4}_{-2.6}$ & $51.4^{+9.8}_{-23.6}$ & $50.0^{+25.0}_{-32.2}$ & - \\
\noalign{\smallskip}\hline\noalign{\smallskip}
\hline
  \end{tabular}
\begin{tablenotes}
\item [$\diamond$] Peculiar SNe~Ia include 91bg-like, SNe~Iax, super-Ch, 00cx-like and Ca-rich SNe~Ia.
\item [$\ast$] Super-Ch candidates by \citet{Scalzo12} not included.
\item [\dag] This cut is applied to objects with better overall 91bg template fits and it requires enough data coverage in blue and red bands. Objects passing it are 91bg-like candidates and the others are other peculiar SNe~Ia.
\end{tablenotes}
\end{threeparttable}
\end{table*}

\subsection{Comparison to other typing techniques}\label{comptype}

Photometric SN classification is an active area of research. Many early studies, ultimately aimed at cosmology, have focussed on normal SN~Ia identification either for prioritization in spectral follow-up \citep{Dahlen02,Sullivan06a} or posterior identification \citep[e.g][]{Riess04b,Barris06} of real datasets, even in the absence of spectroscopic confirmation \citep{Bazin11,Sako11b,Perrett12,Olmstead14}, or of simulated samples \citep{Gong10,Gjergo13}. These techniques most often use model template fits to the light-curve but also color-color diagrams, as well as host galaxy redshift prior information. The study of multiple SN types, including CC-SN classification, is more challenging given the wide diversity of light-curve behavior compared to normal SNe Ia. Building on the early efforts by \citet{Pskovskii78,Pskovskii84}, more recent methods are: color-color evolution \citep{Poznanski02}, color-magnitude evolution \citep{Johnson06}, Bayesian template fitting \citep{Poznanski07}, Bayesian classification schemes \citep{Kuznetsova07}, Fuzzy Set Theory algorithms \citep{Rodney09}, boosting and kernel density estimation techniques \citep{Newling11}, kernel Principal Component Analsys \citep{Ishida13}, a semi-supervised learning \citep{Richards11} and neural networks \citep{Karpenka13}.  

\begin{table}[htbp]
 \centering
\caption{Comparison of normal SN~Ia photometric typing for some of the best selected published techniques in percentages}
 \label{comptable}
  \begin{tabular}{cccc}
 \hline
 \hline
  \textbf{Technique} & \textbf{Efficiency} & \textbf{Purity} & \textbf{FoM} \\
   & ($\%$) & ($\%$) & ($\%$) \\
  \hline
  \citet{Poznanski07} & 97 & 77-91 & 75-88 \\ 
  \citet{Rodney09} & 94 & 98 & 92 \\
  \citet{Sako11b} & 88-92 & 87-94 & 82-86 \\
  \citet{Olmstead14} & 85-92 & 89-93 & 79-83 \\
  {\bf This work} & {\bf 82} & {\bf 99} & {\bf 81} \\
  \hline
 \end{tabular}
\end{table}

A direct comparison to these methods is difficult given the diversity of training samples (although see \citealt{Kessler10b}). Nevertheless, we quote some of these studies final efficiencies, purities and Figures of Merit for normal SN~Ia classification in table~\ref{comptable}. We caution that, unlike our current study, most of these studies have samples with a wide redshift range, where factors like signal-to-noise become important. It is evident that the normal SN~Ia classification is quite robust for all different techniques providing clean samples, $P>90-95$\% (see also \citealt{Gjergo13,Ishida13,Karpenka13}). More challenging is the identification of peculiar groups of SNe~Ia, which are more easily mistaken with CC-SNe, and the present work is a first step in that direction.

\subsection{Photometric ``dromedary'' vs ``camel'' SN~Ia class}


A large range of spectroscopic peculiar SNe~Ia are found here to have many similarities in their light-curve behaviour, resembling those of typical 91bg-like SNe~Ia. SNe~Iax and super-Chandrasekhar SNe~Ia are better fit with a 91bg-like template with SiFTO. This trend is stronger in the redder bands, where post-maximum shoulders and secondary maxima are absent for all of these transients. SN~2006bt-like objects also resemble these, having better red-band 91bg template fits. Grouping them all together, we dub them the photometric ``dromedary'' class, as opposed to classical normal SNe~Ia or the photometric ``camel'' class. For all of these ``dromedary'' objects, SNe~Iax, 06bt-like and super-Chandra, a progressive increase in fit quality (decrease in $\chi^2_{\nu}$) is seen as one moves from short towards longer wavelengths when fitting the light-curves with a 91bg template. The opposite happens when fitting with the normal SN~Ia template. If one were to order these groups according to light-curve similarities with SiFTO fit quality, one would have: 91bg-like $\rightarrow$ Iax $\rightarrow$ super-Chandra $\rightarrow$ 06bt-like $\rightarrow$ normal SNe~Ia.

\begin{figure*}
\centering
\includegraphics[width=1\linewidth]{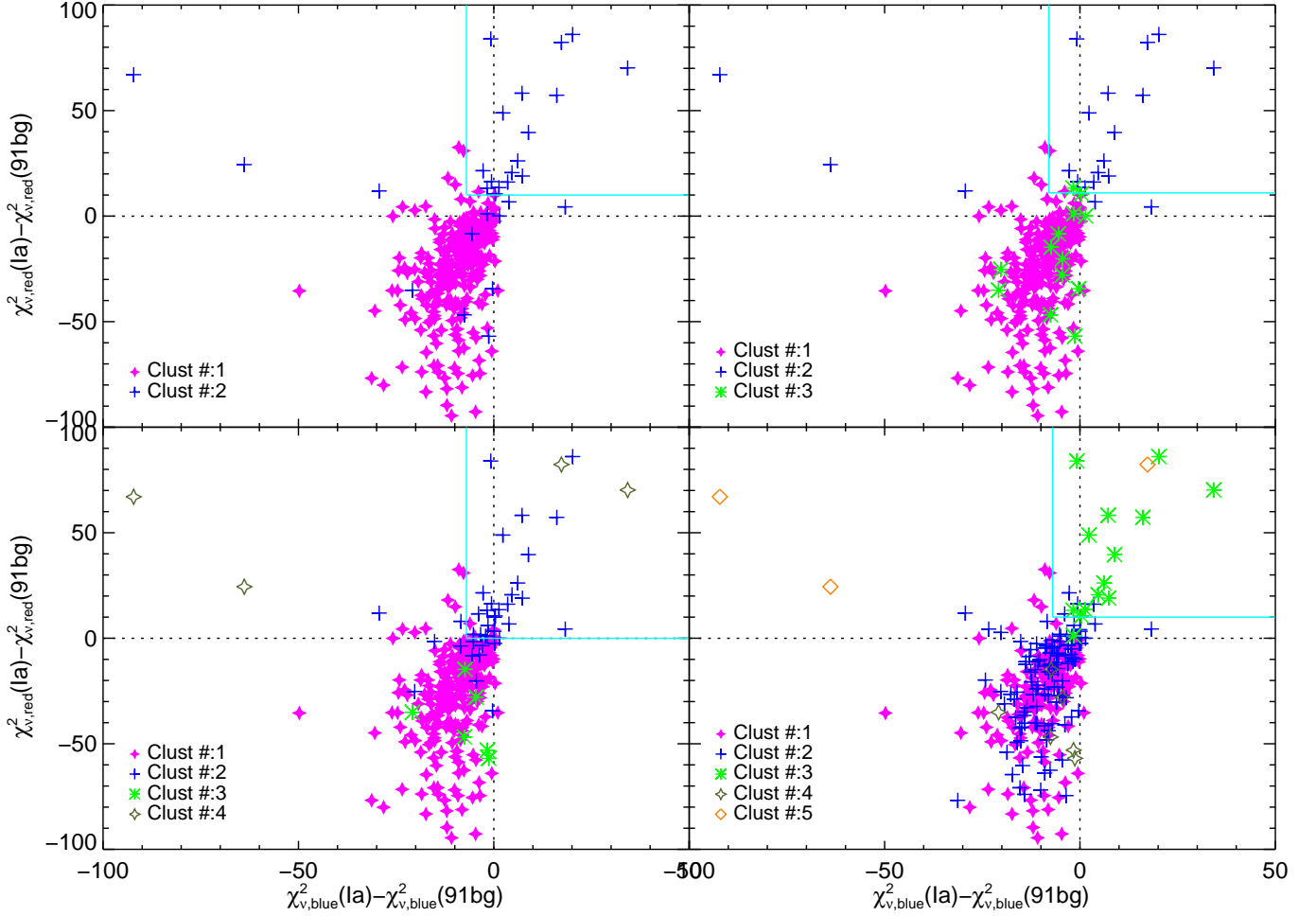}
\caption{Difference between normal and 91bg template fit qualities in the blue bands: $\chi^2_{\nu,\mathrm{blue}}(\mathrm{Ia})-\chi^2_{\nu,\mathrm{blue}}(\mathrm{91bg})$, versus the difference between normal and 91bg template fit qualities in the red bands: $\chi^2_{\nu,\mathrm{red}}(\mathrm{Ia})-\chi^2_{\nu,\mathrm{red}}(\mathrm{91bg})$ (similar to figure~\ref{brchisq}) for different cluster analysis with 2 (top left) to 5 (bottom right) cluster groups. The cluster analysis is based on standardized SiFTO stretch, color, absolute magnitude and overall, blue and red $\chi^2_{\nu}$ for both normal and 91bg templates. Different symbols and colors indicate the groups according to the cluster analysis.  Zero $\chi^2_{\nu}$ difference dotted lines are shown, and solid cyan lines show the region that optimizes the number of objects of the cluster group in the top right corner of each plot (basically 91bg-like SNe~Ia) without objects from other groups.}
       \label{cluster}
\end{figure*}

Furthermore, one can see that the objects of this photometric ``dromedary'' class (excluding 06bt-like SNe), besides being different from the normal SNe~Ia according to two template fits, have different color-stretch (fig.~\ref{colst}) and magnitude-stretch (fig.~\ref{magstcol}a) relations than normal SNe~Ia. In particular they are consistently redder and fainter, i.e. for a particular stretch, different colors and magnitudes are predicted for a normal ``camel'' SN~Ia and a ``dromedary'' SN. This photometric closeness opens up the question if such a variety of SNe~Ia is actually linked to some common physical process, if they arise from similar explosion mechanisms and even if their progenitors are connected. We examine this further in next sections.  

\subsubsection{Cluster Analysis}\label{sec:cluster}

To investigate the hypothesis of physical commonality of SNe in the ``dromedary'' class, we perform a cluster analysis \citep[$k$-means clustering][]{Everitt93} in which, given certain SN characteristics, groups of objects more similar to each other than to the ones of other groups are searched. We provide standardized SiFTO stretch, color, absolute magnitude and $\chi^2_{\nu}$ for overall, blue and red-band fits with both normal and 91bg templates for a total of 12 variables for each SN. We present the results of the cluster analysis in figure~\ref{cluster} requiring $N_{\mathrm{clust}}=2-5$ different cluster groups (from top left to bottom right). No matter the number of cluster groups, in each case we observe clear separations between standard SNe~Ia and 91bg-like SNe~Ia in the top right area of each plot. Most of these SNe in the first quadrant are, as shown in figure~\ref{brchisq}, 91bg-like objects. However, we find that most SNe~Iax also make part of this cluster group ($\sim$5 of 8) strengthening the bond between the two peculiar sets. Super-Chandrasekhar and 06bt-like SNe~Ia, on the contrary, are always grouped together with normal SNe~Ia. This indicates that, even though their light-curves have similarities to 91bg-like objects in the redder bands, overall they are more akin to normal SNe~Ia, as one would expect. We note that in the first $N_{\mathrm{clust}}=2$ cluster analysis, 5 quite normal SNe~Ia are in the 91bg-like group, this is due to their extreme red colors, e.~g. SN~2006X and SN~2003cg, and are put in a different group once we add more cluster groups (or if we do not include color as an input variable). This shows that their colors are of a different nature (interstellar or circumstellar rather than intrinsic) and that the cluster analysis can separate these effects. With $N_{\mathrm{clust}}>3$, the peculiar SN~2005hk, SN~2002es and some extreme cool 91bg-like are grouped together since they lie off the bulk of the population.

For each cluster analysis, we show the box that separates best the objects in the cluster group similar to 91bg-like from the rest of the groups (cyan solid lines). This is done similarly to section~\ref{bluered} by maximizing the FoM of this cluster group. We find boxes ($\Delta\chi^2_{\nu,\mathrm{red}}>10,11,0,10$ and $\Delta\chi^2_{\nu,\mathrm{blue}}>-7,-8,-7,-7$ for each $N_{\mathrm{clust}}=2,3,4,5$ respectively) that compare very well with the box in figure~\ref{brchisq} based on the FoM of known 91bg-objects. This is a strong confirmation of the validity of our technique with an independent and robust appraoch from data mining. Furthermore, the appearance of this separate region in all cluster analysis strongly supports the hypothesis of clear light-curve differences between two SN~Ia populations: normal and 91bg-like/Iax SNe. To check for dependence on the sample, we again perform a bootstrap analysis, similar to section~\ref{bootstrap}, finding full consistency with our results. We obtained following 91bg-like boxes based on the mean and standard deviation of all 200 realizations: $\Delta\chi^2_{\nu,\mathrm{red}}>4.8\pm8.9,1.2\pm8.0,0.2\pm9.6,0.6\pm9.3$ and $\Delta\chi^2_{\nu,\mathrm{blue}}>-10.2\pm6.6,-8.6\pm7.4,-8.3\pm7.9,-7.9\pm8.0$ for each $N_{\mathrm{clust}}=2,3,4,5$ respectively.

\begin{figure*}
\centering
\includegraphics[width=0.7\linewidth]{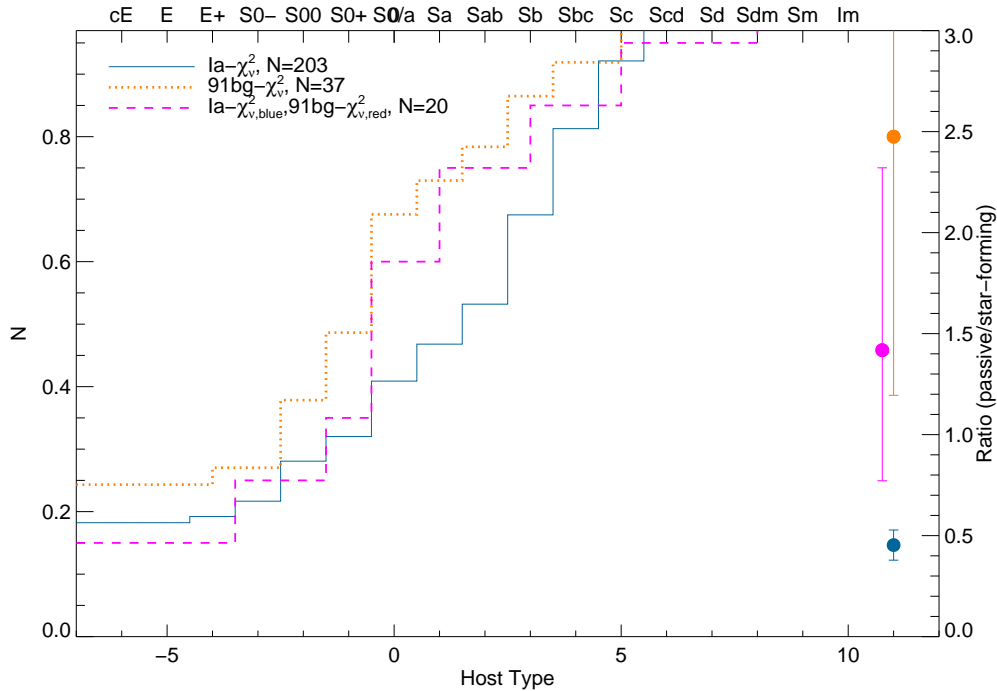}
\caption{Cumulative distributions of morphological SN host T-type from -5 to 10, where T-types below 0 are passive E/S0 galaxies and types greater than 0 are star forming galaxies. Different photometric SN~Ia groups are shown: normal candidates according to overall, blue- and red-band fits (solid blue), the 91bg-like candidates according to overall, blue- and red-band fits (dotted orange) and objects with 91bg-like overall but blue- or red-band normal fits like SNe~Iax and super-Chandrasekhar candidates (dashed purple). In the right column, circles show the ratio of passive to star forming galaxies for the three populations with propagated Poisson errors \citep{Gehrels86}.}
       \label{hosttype}
\end{figure*}

\subsubsection{Environments and progenitors}

From another persepctive, the environments of different SNe~Ia tell us important clues about their progenitors. Although classical 91bg-like objects are particuarly inclined to happen in elliptical and passive star-forming environments \citep[e.~g.][]{Howell01a,Gonzalez11}, PTF~10ops occured far away from its host \citep{Maguire11}, SNe~Iax seem to prefer late-type galaxies \citep{Lyman13,Foley13x} and super-Chandrasekhar appear in various environments with possible preference for low-mass galaxies \citep{Taubenberger11}. In figure~\ref{hosttype}, we show the morphological type of a large subset of our sample binned in T-types from -5 to 10, from E/S0 galaxies to spiral and irregular galaxies. Performing a Kolmogorov-Smirnov (KS) test shows that the normal SN~Ia distribution (solid blue) is significantly different from the distribution of photometric 91bg-like candidates according to overall, blue/red band fits (dotted orange) with a probability P(KS)$\simeq0.002$. On the other hand, other peculiar SNe~Ia according to the overall but not blue or red-band fits, i.~e. objects such as PTF~10ops-like, SNe~Iax and super-Chandra together (dashed purple) are not statistically different from normal SNe~Ia (P(KS)$\simeq0.12$), nor from 91bg-like candidates (P(KS)$\simeq0.92$). In the right bin of the figure we show that the ratio of passive to star forming hosts for the photometric normal SN~Ia population (blue circle) is much lower than for the photometric 91bg-like candidates (orange cirle), which is a well-known relation. Interestingly, the ratio of the other peculiar SNe~Ia (purple circle) lies in between those two samples. A Fisher exact test \citep{Fisher22} gives a similar result to the KS test: the number ratio of photometric normal SNe~Ia in passive to star-forming galaxies is statistically different from the ratio of photometric 91bg-like SNe~Ia with a probability P(F)$\simeq0.002$, whereas the ratio for peculiar SNe~Ia in passive to star-forming galaxies is not statistically different from the ratio for normal SNe~Ia (P(F)$\simeq0.21$), nor from the ratio of 91bg-like SNe~Ia (P(F)$\simeq0.50$). Albeit with low statistics, this may point towards the peculiar SN~Ia group being transitional in progenitor characteristics between normal and 91bg-like, or more likely, it evidences the presence of a mixed group with transitional objects, some 91bg-like and some normal SNe~Ia. 

From a theoretical standpoint, SNe~Iax, SN~1991T and super-Chandra SNe~Ia all have hot, highly ionized photospheres as has been shown spectroscopically \citep{Foley13x}. In the scenario of \citet{Kasen06} and \citet{Kasen07}, the secondary maximum at long wavelengths can be a direct consequence of the iron group abundance stratification: if the SN has the iron concentrated in the inner region, the recombination of doubly ionized Fe and Co into singly ionized will only happen at later times, redistributing radiation from the blue to the red and creating the secondary maximum. The absence of secondary maxima in red bands for these hot objects can thus be explained with high mixing of $^{56}$Ni in the outer layers \citep[see also][]{Scalzo12} and argues against a delayed detonation mechanism. In this same framework, the absence of secondary maxima in 91bg-like SNe~Ia is explained with quite an opposite argument: these are very cool objects and the recombination therefore sets on much earlier making the first maximum coincide with the second. So, in one case it's mixing and in the other temperature that would explain the post-maximum red band behaviour. Although 91bg-like objects are spectroscopically different to the other peculiar sub-classes, there are some intriguing transitional objects between 91bg-like and Iax with spectroscopic similarities \citep{Ganeshalingam12} that may prove a continous set of explosion mechanisms.

Regarding the progenitor systems, the differences among these peculiar SNe~Ia are strengthened. Linking classical 91bg-like and high-stretch PTF~10ops-like SN~Ia progenitors, \citet{Taubenberger13} argue that the nebular spectra of the high-stretch 91bg-like SN~2010lp presents clear evidence for a low density in the core that is only consistent with a violent merger of two CO-WDs \citep{Kromer13}. \citet{Pakmor13}, on the other hand, show that 91bg-like explosions can be reproduced via violent mergers of a CO-WD and a He-WD. Super-Chandrasekhar explosions in turn typically invoke slow double-degenerate mergers of massive CO-WDs \citep[e.~g.][]{Hachinger12}. These all invoke variate scenarios of double-degenerate systems. On the contrary, the most favoured SN~Iax progenitor scenarios are single-degenerate in origin: double detonations of a CO-WD and a He-star \citep{Foley13x,Wang13} or pure deflagrations of Chandrasekhar mass WDs \citep{McClelland10,Jordan12,Kromer13x}. As for normal SNe~Ia, another mechanism such as the prototypical single-degenerate delayed detonation \citep[e.~g.][]{Blondin13,Sim13} or even double-degenerate mergers with higher WD masses \citep{Pakmor13} could explain the differences with the others SNe. Sub-Chandrasekhar explosions have also gained renewed popularity through the double detonation scenario arising from a shell detonation \citep{Woosley86,Bildsten07,Waldman11,Fink10}, or via violent mergers of two WDs \citep{Pakmor12}, or even through WD collisions \citep{Benz89,Rosswog09,Raskin09}.

Putting this information together, although photometrically, the ``dromedary'' SN~Ia class presents interesting common properties among its variate members, such as red color and lack of secondary maximima at longer wavelengths, a cluster analysis suggests that super-Chandrasekhar and 06bt-like are more similar to normal SNe~Ia. SNe~Iax and 91bg-like SNe are similar photometrically between each other, but they are known to have spectroscopic features that differ as do their environments. None of the members of the dromedary class can be well explained with a delayed detonation explosion, yet no other clear theoretical link exist among all members of this class. We therefore conclude that although there probably is some common physical mechanism driving the similar behaviour in the cooler redder part of their emission at epochs after maximum, their progenitor and explosion are not necessarily similar in nature. 

\subsection{Recipe to photometrically identify different SN~Ia sub-groups}
In this paper we have presented a photometric algorithm to identify normal SNe~Ia and 91bg-like SNe~Ia, the classical fast ones but also the new class of slowly declining 91bg-like SNe~Ia such as PTF~10ops and SN~2010lp. The method also allows us to identify numerous peculiar SNe~Ia, particularly SNe~Iax, super-Chandrasekhar SNe~Ia and SN~2006bt-like objects. For it to work at best, we recommend photometry before and after maximum light in several bands, in at least one blue and one red band. At higher distances, where the emission is redshifted, these bands will need to be adjusted. Here is a step-by-step recipe to classify the various events:

\begin{itemize}
\item {\bf SiFTO fits:} Fit each SN with SiFTO using the two templates, normal and 91bg. Also fit each SN with the two templates using only blue-bands and only red-bands.
\item {\bf CC contamination:} Use $\chi^2_{\nu}$ cuts to remove CC contamination. For our bootstrapped sample: $\chi^2_{\nu}(\mathrm{Ia})<34.0\pm6.7$ and $\chi^2_{\nu}(\mathrm{91bg})<21.7\pm4.0$ keeps SNe~Ia.
\item {\bf Normal SNe~Ia:} Objects passing the previous cut and with $\chi^2_{\nu}(\mathrm{Ia})-\chi^2_{\nu}(\mathrm{91bg})<0$ are normal SN~Ia candidates with high confidence. One can reduce the CC contamination of all SNe~Ia further with cuts in color-color diagrams. We use $(u-B)_{10}>(0.18\pm0.13)$ and $(B-V)_{10}<0.7(u-B)_{10}+(0.77\pm0.23)$. The selected sample may have a very small fraction of objects being of CC.
\item {\bf 91bg-like SNe~Ia:} Objects that have better overall, blue-band and red-band 91bg template fits (with a definition of $\chi^2_{\nu,\mathrm{blue,red}}(\mathrm{Ia})-\chi^2_{\nu,\mathrm{blue,red}}(\mathrm{91bg})>(-0.8\pm2.2),(2.8\pm3.1)$) are considered 91bg-like SN~Ia candidates. If they have low-stretch, i.e. $s_{\mathrm{91bg}}<1.1$, then they are the classical 91bg-like objects, at higher stretch (and $\mathcal{C}>0$), they are like PTF~10ops or SN~2010lp.
\item {\bf Super Chandrasekhar SNe~Ia:} Those objects with better overall 91bg fits but better normal blue or red-bands fits, that also have very wide light-curves, i.e. $s_{\mathrm{Ia}}>1.2$ (or $s_{\mathrm{91bg}}>1.8$), are typical super-Chandrasekhar candidates.
\item {\bf SNe~Iax:} SNe~Ia with better overall 91bg fits but better normal blue- or red-bands fits, and narrower light-curves than in the previous point are probable SNe~Iax. There is a possibility that some of these are objects like SN~2002es, of unkown origin or transitional objects, i.e. maybe 91bg-like, maybe SN~2006bt-like, in which case there is no way to disentangle between those two groups according to our method. Given the numbers, it is more probable ($\sim83\%$ with the current sample) that such an SN will be a type Iax SN.
\item {\bf SN~2000cx-like SNe~Ia:} If the SN~Ia has better overall 91bg but worse red- or blue-band fit, \emph{and} it additionally has a blue color, $\mathcal{C}<0$, it is most probably not a 91bg-like high-stretch SN~Ia nor a SN~Iax but a peculiar SN~2000cx object, or rather a SN~2013bh-like object.
\item{\bf SN~2006bt-like SNe~Ia:} Finally, objects with better overall normal template fit but better red-band 91bg fits, and wider light-curves ($s_{\mathrm{91bg}}>1.3$) are possibly a group similar to SN~2006bt. These could be a link between normal and super-Chandrasekhar SNe~Ia.
\end{itemize}

\section{Summary}\label{summary}
We have investigated the light-curves of a large sample of low-$z$ SNe~Ia with the light-curve fitter SiFTO and two spectral template series for normal and for SN~1991bg-like SNe~Ia. By comparing the fit with the two templates in different filter sets, we are able to photometrically identify typical 91bg-like SNe~Ia and also those spectroscopically similar but with wider light-curves such as recent PTF~10ops and SN~2010lp. We confirm the robustness of the technique comparing our photometric typing technique with different spectroscopic classificators such as SNID. We find that for standard fast-evolving 91bg-like SNe~Ia the stretch obtained with a 91bg template fit, $s_{\mathrm{91bg}}$, is better at describing this sub-group than a normal template stretch, $s_{\mathrm{Ia}}$. The existence of two groups is strengthened by a cluster analysis that suggests two different populations based on a set of SN photometric properties. Despite this, we point out to the smooth transition between the two and the existence of transitional objects between normal and 91bg-like SNe. Regarding PTF~10ops-like transients, i.e. 91bg-like of wide light-curves, we find 3-4 possible candidates in the literature, confirming the rarity or different environment of such events.

Furthermore, we find a range of transient light-curves like SNe~Iax, super-Chandrasekhar SNe~Ia and SN~2006bt-like that are more similar to a 91bg-like than a normal template to varying degrees at longer wavelengths. All these ``dromedary'' objects lack the characteristic prominent secondary maxima or shoulders seen in red filters of classical SNe~Ia and could suggest similar physical processes. Using fit qualities in the different filter sets, we are able to disentangle most of them from typical 91bg-like (which have better 91bg template fits in both blue and red filter sets) and normal SNe~Ia (which have worse 91bg template fits in all filter sets). Of these peculiar objects, SNe~Iax generally resemble 91bg-like objects more closely, also in their absolute magnitudes, and some of them can therefore not be differentiated with our technique. The cluster analysis also joins most of SNe~Iax with the 91bg-like group, showing that without spectroscopic properties, the light-curve differences are not strong enough to form two different groups. Super-Chandra SNe~Ia, on the other hand, are clearly distinct in several regards from other peculiar SNe~Ia resembling more normal SNe~Ia, and it is easy to select them out. We propose an interesting relation between super-Chandra and SN~2006bt-like SNe~Ia, with the latter being a less extreme case of the most luminous of SNe~Ia, and possibly even also a link with SNe~Iax and slowly declining 91bg-like objects.

We have presented a simple yet powerful technique to identify different sub-groups of SNe~Ia in a purely photometric manner. With a competitive FoM(Ia)$>80\%$, we have shown that it robustly classifies normal SNe~Ia with very little contamination from CC~SNe, but also from other peculiar SNe~Ia that are unsuit for cosmological studies. Alternatively, it also allows to search for these peculiar SNe~Ia to investigate them further: 91bg-like objects, SNe~Iax and even super-Chandrasekhar SNe. Thus, this method is applicable for photometric classification of transients in coming wide field large surveys such as LSST, and to prioritize the study of these interesting peculiar objects, or, on the other hand, to reject them for studies of purely normal SNe~Ia.

\vspace{10mm}
\section*{Acknowledgements}
We thank the anonymous referee for useful comments that improved this work. This paper has made use of a large public data set coming from the long lasting effort of a variety of surveys and groups, for which we are extremely thankful. The work of CSP has been supported by the National Science Foundation under grants AST~0306969, AST~0607438 and AST~1008343. The CfA Supernova Archive is funded in part by the National Science Foundation through grant AST~0907903. This research has also made use of the NASA/IPAC Extragalactic Database (NED) which is operated by the Jet Propulsion Laboratory, California Institute of Technology, under contract with the National Aeronautics. 

S.G., F.B. and L.G. acknowledge support from CONICYT through FONDECYT grants 3130680, 3120227 and 3140566, respectively. Support for S.G., G.P., F.F, C.G, F.B., L.G., M.H. and T.J. is provided by the Ministry of Economy, Development, and Tourism's Millennium Science Initiative through grant IC12009, awarded to The Millennium Institute of Astrophysics, MAS.
\vspace{5mm}

\bibliographystyle{apj}
\bibliography{astro}

{\centering
\begin{longtable*}[ht]{ccccp{10cm}}
 \caption{Nearby SNe~Ia used in this study requiring enough light-curve coverage. Reduced quality of fit ($\chi^2_{\nu}$) to a normal and a 91bg template are shown.}\label{Iaphot} \\
    \hline \hline \vspace{0.05in} \\
    \textbf{Name} &   {\textbf{Type}} & \boldmath{$\chi^2_{\nu}$}\textbf{(Ia)} &  \boldmath{$\chi^2_{\nu}$}\textbf{(91bg)} & \textbf{Source}  \vspace{0.05in} \\
    \hline \vspace{0.05in} \\
    \endfirsthead
    \multicolumn{5}{c}{{\tablename} \thetable{} -- {\textbf{Continued}}} \\[0.5ex]
    \hline \hline \vspace{0.05in} \\
    {\textbf{Name}} &  {\textbf{Type}} & \boldmath{$\chi^2_{\nu}$}\textbf{(Ia)} &  \boldmath{$\chi^2_{\nu}$}\textbf{(91bg)} & \textbf{Source}  \vspace{0.05in} \\
    \hline \vspace{0.05in} \\
    \endhead

    PTF09dav  &    SN~Ia-Ca & 8.20  &     3.86  &                                                   \citet{Sullivan11a} \\
    PTF10bjs  &    SN~Ia & 5.86  &    20.13  &                                                          \citet{Hicken12} \\
    PTF10ops  &    SN~Ia & 9.99  &     2.08  &                                                     \citet{Maguire11} \\
     SN~1980N  &   SN~Ia &  5.19  &     8.16  &                                                          \citet{Hamuy91} \\
     SN~1981B  &   SN~Ia & 15.24  &    26.41  &                   \citet{Barbon82} \\
     SN~1981D  &   SN~Ia &  0.84  &     1.28  &                                                          \citet{Hamuy91} \\
     SN~1982W  &   SN~Ia & 574.22  &   437.40  &                                                                          \citet{Barbon89} \\
     SN~1983G  &   SN~Ia & 32.59  &    36.41  &                \citet{Younger85,Benetti91,Cadonau90} \\
\end{longtable*}
}

{\centering
\begin{longtable*}[ht]{ccccp{10cm}}
 \caption{Nearby CC~SNe used in this study requiring enough light-curve coverage. Reduced quality of fit ($\chi^2_{\nu}$) to a normal and a 91bg-like SN~Ia template are shown.}\label{CCphot} \\
    \hline \hline \vspace{0.05in} \\
    \textbf{Name} & \textbf{Type} & \boldmath{$\chi^2_{\nu}$}\textbf{(Ia)} &  \boldmath{$\chi^2_{\nu}$}\textbf{(91bg)} & \textbf{Source}  \vspace{0.05in} \\
    \hline \vspace{0.05in} \\
    \endfirsthead
    \multicolumn{4}{c}{{\tablename} \thetable{} -- {\textbf{Continued}}} \\[0.5ex]
    \hline \hline \vspace{0.05in} \\
    {\textbf{Name}} & \textbf{Type} & \boldmath{$\chi^2_{\nu}$}\textbf{(Ia)} &  \boldmath{$\chi^2_{\nu}$}\textbf{(91bg)} & \textbf{Source}  \vspace{0.05in} \\
    \hline \vspace{0.05in} \\
    \endhead

     SN~1986I  &       SN~II  &    14.70  &    12.72  &                                                          \citet{Pennypacker89} \\
     SN~1987A  &       SN~II  &   252.44  &   282.08  & \citet{Hamuy90,Bouchet89} \\
     SN~1988A  &       SN~II  &     3.35  &     2.40  &                         \citet{Benetti91,Turatto93} \\
     SN~1988H  &       SN~II  &     0.15  &     0.15  &                                                          \citet{Turatto93} \\
     SN~1990E  &       SN~II  &    41.89  &    20.40  &                           \citet{Benetti94,Schmidt93} \\
     SN~1990K  &       SN~II  &    23.81  &    23.31  &                                                              \citet{Cappellaro95} \\
     SN~1991G  &       SN~II  &   100.56  &   126.02  &                                                            \citet{Blanton95} \\
    SN~1992am  &       SN~II  &   145.46  &   126.15  &                    \citet{Schmidt94} \\
\end{longtable*}
}

\end{document}